\documentclass[11pt,english]{revtex4-2}
\usepackage[LGR,T1]{fontenc}
\usepackage[latin9]{inputenc}
\usepackage{geometry}
\usepackage{amsmath}
\usepackage{amssymb}
\usepackage{graphicx}
\usepackage{esint}

\makeatletter

\DeclareRobustCommand{\greektext}{%
  \fontencoding{LGR}\selectfont\def\encodingdefault{LGR}}
\DeclareRobustCommand{\textgreek}[1]{\leavevmode{\greektext #1}}
\ProvideTextCommand{\~}{LGR}[1]{\char126#1}

\makeatother

\usepackage{babel}
\begin{document}
\global\long\def\ket#1{\left|#1\right\rangle }%

\global\long\def\bra#1{\left\langle #1\right|}%

\global\long\def\ketL#1{\left.\left|#1\right\rangle \right\rangle }%

\global\long\def\braL#1{\left\langle \left\langle #1\right|\right.}%

\global\long\def\braket#1#2{\left\langle #1\left|#2\right.\right\rangle }%

\global\long\def\ketbra#1#2{\left|#1\right\rangle \left\langle #2\right|}%

\global\long\def\braOket#1#2#3{\left\langle #1\left|#2\right|#3\right\rangle }%

\global\long\def\mc#1{\mathcal{#1}}%

\global\long\def\nrm#1{\left\Vert #1\right\Vert }%

\title{Pseudo Twirling Mitigation of Coherent Errors in non-Clifford Gates}
\begin{abstract}
The conventional circuit paradigm, utilizing a limited number of gates
to construct arbitrary quantum circuits, is hindered by significant
noise overhead. For instance, the standard gate paradigm employs two
CNOT gates for the partial CPhase rotation in the quantum Fourier
transform, even when the rotation angle is very small. In contrast,
some quantum computer platforms can directly implement such operations
using their native interaction, resulting in considerably shorter
and less noisy implementations for small rotation angles. Unfortunately,
coherent errors stemming from qubit crosstalk and calibration imperfections
render these implementations impractical. In Clifford gates such as
the CNOT, these errors can be addressed through Pauli twirling (also
known as randomized compiling). However, this technique is not applicable
to the non-Clifford native implementations described above. The present
work introduces, analyzes, and experimentally demonstrates a technique
called `Pseudo Twirling' to address coherent errors in general gates
and circuits. Additionally, we experimentally showcase that integrating
pseudo twirling with a quantum error mitigation method called `Adaptive
KIK' enables the simultaneous mitigation of both noise and coherent
errors in non-Clifford gates. Due to its unique features pseudo twirling
could become a valuable asset in enhancing the capabilities of both
present and future NISQ devices.
\end{abstract}
\author{Jader P. Santos}
\author{Ben Bar}
\author{Raam Uzdin}
\affiliation{Fritz Haber Research Center for Molecular Dynamics, Institute of Chemistry,
The Hebrew University of Jerusalem, Jerusalem 9190401, Israel}
\maketitle

\section{Introduction}

In recent years, quantum error mitigation protocols (QEM) \citep{cai2022quantum,endo2018practical,suzuki2022quantum,temme2017error,Huggins2021,Koczor2021,PhysRevA.102.012426,strikis2021learning,li2017efficient,TEMqem}
have significantly boosted the performance of quantum computers \citep{kim2023evidence,kandala2019error,Huggins2021,song2019quantum,zhang2020error,PhysRevX.11.041039,Plenio2023noiseAssistedIBM,sagastizabal2019experimental,van2023probabilistic,shtanko2023uncovering}.
These methods are primarily designed for managing incoherent errors
(noise) arising from interactions with the environment. However, in
addition to incoherent errors, there are also coherent errors that
cannot be addressed using the same methods in a scalable manner. Coherent
errors pose a significant concern in the development of quantum computers
in the foreseeable future. They are often attributed to calibration
errors, drift in calibration parameters, or coherent crosstalk interactions
between qubits. While hardware continues to advance, the demand for
minimizing coherent errors becomes increasingly crucial as circuits
grow in size and complexity.

Currently, the most effective approach to address these errors is
a method known as Randomized Compiling (RC) \citep{wallman2016noise,PhysRevX.11.041039,Emerson2020LearnTwirlExpIBM,knill2004PauliTwirl,RCqutrits},
also referred to as Pauli twirling. However, this technique is limited
to Clifford gates. For Clifford gates such as the CNOT, iSWAP, and
others, RC involves executing an ensemble of circuits that are equivalent
in the absence of coherent errors, but each circuit alters the coherent
error in a distinct way. After running these circuits and performing
post-processing averaging, the coherent error is mitigated, but the
resulting transformation is no longer a unitary. In other words, the
coherent error is converted into an effective incoherent error. This
conversion is beneficial as incoherent errors can be addressed with
QEM protocols. As demonstrated experimentally in \citep{KIKarxiv}
the integration of RC with a QEM method called `Adaptive KIK', was
imperative for removing coherent errors and obtain accurate results.

Another interesting feature of RC is that it converts a general incoherent
error channel into a Pauli channel that has a much simpler structure
\citep{wallman2016noise}. Due to the reduced degrees of freedom in
the Pauli channel compared to general noise, it becomes feasible to
employ additional sparsity techniques and efficiently characterize
the noise channel of multi-qubit Clifford gates \citep{van2023probabilistic,2023EfficientPauliLearning,Emerson2020LearnTwirlExpIBM}.
This key feature facilitates two important QEM techniques, PEC \citep{temme2017error}
and PEA \citep{kim2023evidence}. PEA presently holds the record of
the largest circuit that QEM has been applied to -- 127 qubit Ising
lattice time-evolution experiment \citep{kim2023evidence}. The largest
depth probed in this experiment was sixty cnots. 

Unfortunately, RC is restricted to Clifford gates, rendering PEC and
PEA inapplicable to non-Clifford multi-qubit gates. A recent work
\citep{laydenPECnonCliff} extends these methods to non-Clifford gates.
In contrast to the Clifford case, the sampling overhead per gate can
be substantial regardless of how small the coherent error is. In \citep{kim2023scalable}
a partial RC was applied by using only the Paulis that commutes with
the non-Clifford gate. This partial twirling can only address coherent
errors that happen to anti-commute with the selected Pauli subset.
Thus, it is not possible to rely on this method if the coherent error
is of unknown type, or if it known to commute with the selected Pauli
subset.

Circuits consisting solely of Clifford gates can be efficiently simulated,
making it pointless to build a powerful quantum computer just for
their execution. Although multi-qubit non-Clifford gates are not part
of the conventional universal gate set paradigm, they can prove crucial
in specific applications, particularly in the era of noisy quantum
computers where circuit depth is a limiting factor.

\begin{figure}
\begin{centering}
\includegraphics[width=12cm]{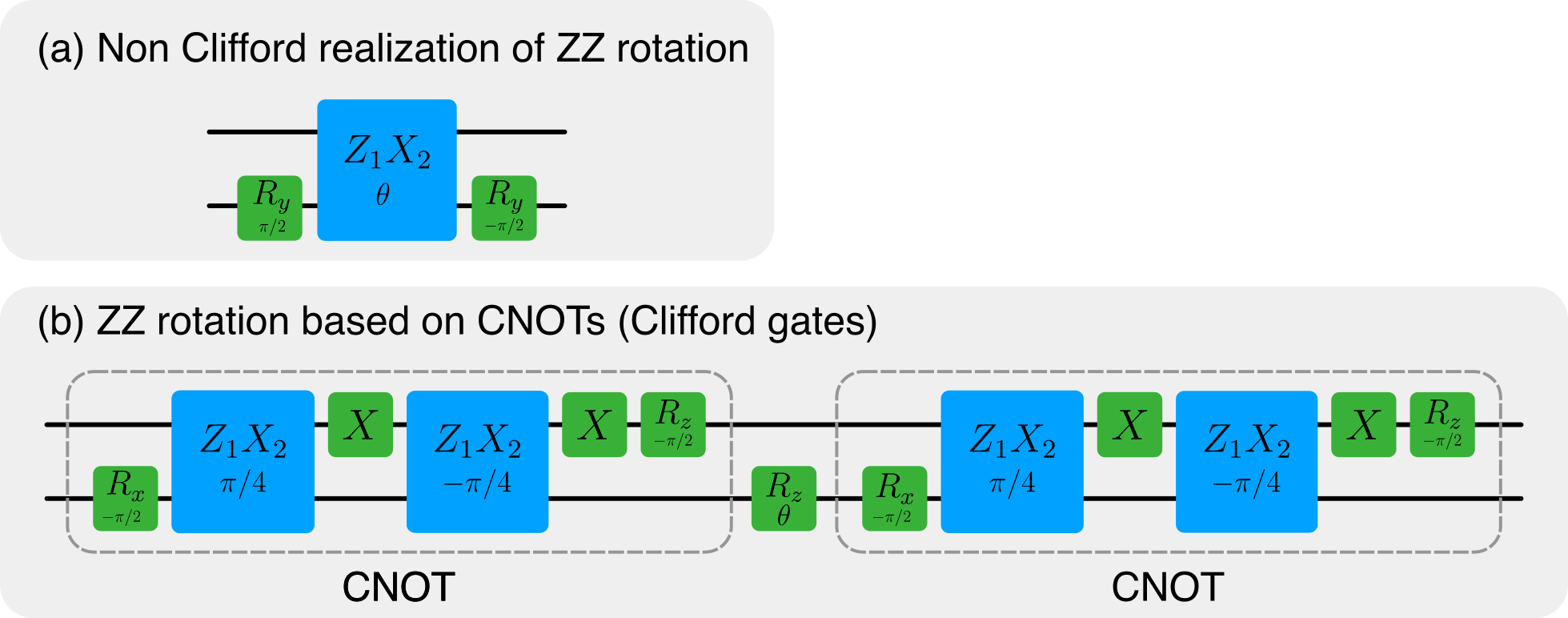}\caption{\label{fig: ZZ}Non-Clifford vs. Clifford implementation of quantum
gates. (a) The useful two-qubit gate $R_{zz}(\theta)=e^{-i\frac{1}{2}\theta Z_{1}Z_{2}}$
can be implemented compactly in the IBM platform using a non-Clifford
gate $R_{zx}(\theta)=e^{-i\frac{1}{2}\theta Z_{1}X_{2}}$, generated
by the cross-resonance interaction. Conversely, the current Clifford-based
implementation (b) is substantially longer, with each box representing
a CNOT gate. Consequently, this implementation exhibits considerably
higher levels of noise. However, currently, there is no technique
for addressing coherent errors in non-Clifford gates, providing an
advantage to the more noisy implementation depicted in (b). In this
work we present a protocol for mitigating coherent errors in compact,
low-noise non-Clifford gates, as shown in (a).}
\par\end{centering}
\end{figure}

As a first example, we bring the time-evolution of the transverse
Ising lattice model \citep{suzuki2012IsingQuantum}. The circuits
implementing this evolution require $R_{zz}(\theta)$ gates that can
be directly implemented very rapidly, leading to significantly lower
noise levels. For example, in \citep{kim2023scalable} the IBM cross-resonance
interaction \citep{McKa2020CRgateFirstP,sundaresan2020CRgateReducing,PulseECRgate},
was facilitated to directly implement $R_{zz}(\theta)$ using two
additional single qubit gates as depicted in Fig. \ref{fig: ZZ}a.
In contrast, in the cNOT gate paradigm, the length of the ZZ rotation
gate is determined by the length of two CNOT gates, even when the
rotation angle is minute (see \ref{fig: ZZ}b). The direct implementation
in \citep{kim2023scalable} has enabled the execution of Ising time
evolution that are four times longer compared to the double-CNOT implementation
(Fig. \ref{fig: ZZ}b). In trapped ions, a similar statement holds
for the Mølmer--Sørensen gate \citep{roos2008ionMSgate}. Although
in this case the direct non-Clifford gate may not be shorter than
a single CNOT gate, it is at least twice shorter than the double-CNOT
implementation.

The second example pertains to the $n$-qubit quantum Fourier transform
\citep{nielsen2002quantum} that comprises of single-qubit gates and
two-qubit gates that execute CPhase($\theta$) rotations with an angle
of $\theta=\pi/2^{n}$ (and multiples of it). As $n$ increases, the
gate performs a transformation that is very close to the identity
transformation. As such, a substantial advantage in terms of noise
is expected when using the direct implementation (the CPhase($\theta$)
analogue of Fig. \ref{fig: ZZ}a) instead of the double-CNOT implementation.
Hence, a method for mitigating coherent errors in non-Clifford gates
would benefit the implementation quantum Fourier transforms and other
circuits as well.

In this work we present a new twirling scheme called `Pseudo Twirling'
(PST). This approach offers three noteworthy characteristics: 1) PST
is equally applicable to both Clifford and non-Clifford gates. 2)
In contrast to RC (Pauli twirling), PST distinguishes between controlled
over(under)-rotation errors and all other controlled or uncontrolled
coherent errors. Over-rotations typically arise from imperfect initial
calibration or a calibration drift. Others errors stem from various
mechanism ranging from crosstalk between qubits to parasitic higher-order
interactions created the driving fields. PST transforms all coherent
error excluding controlled rotation errors - a fact that can be exploited
for high accuracy calibration. 3) PST does not produce a Pauli error
channel; instead, it simplifies the noise in a different manner. In
this paper we present and study these three features. The integration
of PST with QEM, as demonstrated in the present work, can have an
immediate impact on the capabilities of quantum computers. 

With the restriction of applying twirling exclusively to Clifford
gates removed, PST can be employed within individual components of
a gate (intra-gate PST). As we show later, this facilitates more effective
mitigation of coherent errors since coherent errors are addressed
before they have a chance to accumulate. Additionally, PST's flexibility
allows for its application across an entire circuit, where the twirling
occurs at the edges. Although this `Edges PST' is less effective than
intra-gate PST, it can capture coherent errors occurring on a scale
exceeding that of multi-qubit gates.

\subsection*{Notations and quantum dynamics in Liouville space}

For convenience we shall use Liouville space representation where
the density matrix is flattened into a column vector $\ket{\rho}$
and the evolution is described by superoperators operating on this
vector from the left $\ket{\rho(t)}=\mc K(t)\ket{\rho(0)}$. If the
dynamics is unitary the evolution of the density matrix $\ket{\rho(t)}$
is determined by Schrödinger-like equation where
\begin{align}
\frac{d}{dt}\ket{\rho(t)} & =-i\mc H(t)\ket{\rho(t)},\\
\mc H(t) & =H(t)\otimes I-I\otimes H(t)^{\intercal}.\label{eq: Hlio}
\end{align}
where $H(t)$ and $I$ are the Hamiltonian and identity operator in
Hilbert space, and $\intercal$ stands for transposition. The resulting
unitary in Liouville space is
\begin{equation}
\mc U(t)=U(t)\otimes U(t)^{*},\label{eq: Ulio}
\end{equation}
where $U(t)$ is the evolution operator in Hilbert space. Pauli matrices
in Liouville space can either appear as Hamiltonians or as unitaries,
but unlike in Hilbert space, these two forms differ. Let us denote
by $P_{\alpha}$ the tensor product of single-qubit Pauli matrices
$\sigma_{i}$ such that $P_{\alpha}\in\{\sigma_{k}\otimes\sigma_{l}\otimes\sigma_{m}...\}_{k,l,m..\in\{0,x,y,z\}}$.
A Pauli Hamiltonian in Liouville space is given by $\mc H_{\alpha}=P_{\alpha}\otimes I-I\otimes P_{\alpha}^{\intercal}$,
while according to Eq. (\ref{eq: Ulio}), a Pauli evolution operator
has the form $\mathcal{P}_{\alpha}=P_{\alpha}\otimes P_{\alpha}^{*}$
in Liouville space. These two are related via $\mathcal{P}_{\alpha}=exp(-i\frac{\pi}{2}\mc H_{\alpha})$.
See Appendix I for more details. 

\section{Description of the PST methodology }

\subsection{Regular Randomized Compiling}

For clarity, we start by describing the regular randomized compiling/Pauli
twirling scheme, which is applicable only for Clifford gates. Let
$\mc U_{\text{cliff}}$ stand for the evolution operator of an ideal
Clifford gate, and let $\mc K_{\text{cliff}}$ stand for the evolution
operator of the noisy realization of $\mc U_{\text{cliff}}$. $\mc K_{\text{cliff}}$
may encompass both noise and coherent errors, but, for the moment,
we will concentrate on the noise-free scenario where only coherent
errors take place. 

\begin{figure}
\begin{centering}
\includegraphics[width=12cm]{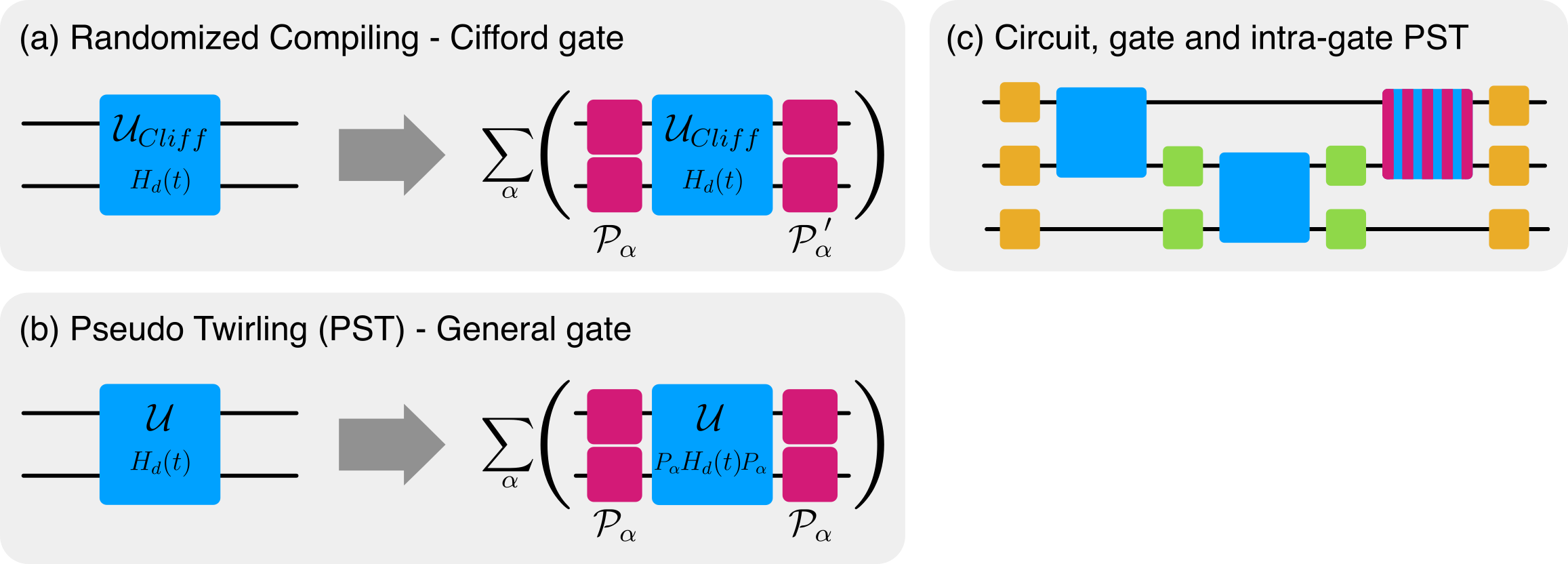}\caption{\label{fig: RC_PST}(a) Randomized compiling (RC, AKA Pauli twirling)
replaces the original Clifford (e.g. a CNOT) with a sum of Pauli-twirled
Clifford gates. RC can only mitigate coherent errors in Clifford gates.
(b) Our proposed Pseudo Twirling (PST) scheme can mitigate uncontrolled
coherent errors, such as crosstalk, in a general gate. This is achieved
by modifying the signs of some of the pulses in the driving field
$H_{d}(t)$ that generates the gate according to the rule $H_{d}(t)\to P_{\alpha}H_{d}(t)P_{\alpha}$.
(c) Because PST is not restricted to Clifford gates, it can be implemented
at various levels of a circuit: at the entire circuit level (orange
lines), gate-level (green lines), and intra-gate level (purple lines). }
\par\end{centering}
\end{figure}

As shown in Fig. \ref{fig: RC_PST}a, in RC, $\mc K_{\text{cliff}}$
is replaced by an averaged operator, with each element ``twirled''
by an n-qubit Pauli operators $\mc P_{\alpha}$ and $\mc P'_{\alpha}$
\begin{equation}
\mc K_{\text{cliff}}^{RC}=\frac{1}{2^{2n}}\sum_{\alpha=1}^{2^{2n}}\mc P_{\alpha}'\mc K_{\text{cliff}}\mc P_{\alpha}.
\end{equation}
$\mc P'_{\alpha}$ is chosen in such a way that it yields $\mc K_{\text{cliff}}^{RC}=\mc U_{\text{cliff}}$
in the absence of coherent and incoherent errors. This condition of
invariance can be expressed as $\mc P_{\alpha}'\mc U_{\text{cliff}}\mc P_{\alpha}=\mc U_{\text{cliff}}$,
leading to the determination of $\mc P_{\alpha}'$ as $\mc P_{\alpha}'=\mc U_{\text{cliff}}\mc P_{\alpha}\mc U_{\text{cliff}}^{\dagger}$.
Therefore, $\mc P_{\alpha}'$ is determined by the selected $\mc P_{\alpha}$.
Up to his point it appears as if we have not used the fact that $\mc U_{\text{cliff}}$
is a Clifford gate. However, since Clifford gates normalize the Pauli
operators, it is guaranteed that if $\mc P_{\alpha}$ is a Pauli operator,
then $\mc P_{\alpha}'$ is also a Pauli operator. Conversely, if the
ideal gate is non-Clifford, $\mc P_{\alpha}'$ may involve multi-qubit
interactions, introducing noise and coherent errors comparable to
those RC aims to correct. These additional errors could potentially
degrade the performance of the gate with respect to the non-twirled
gate. Thus, a significant advantage of Pauli Twirling lies in the
experimental simplicity of the twirling operations, which involve
very low levels of coherent and incoherent errors.

Notably, the total number of twirlings for $n$-qubit gate, $2^{2n}$,
can swiftly grow to a very substantial value. This becomes particularly
evident when considering sequences of twirled gates. Fortunately,
there is no need to sum over all possible Pauli operators. It suffices
to sample from the set $\{\mc P_{\alpha}\}{}_{\alpha=1}^{2^{2n}}$
to obtain the average value. Nevertheless, for the sake of simplicity
of the derivations we will maintain the use of $2^{2n}$.

To gain a better grasp of how RC operates, let us express $\mc K_{\text{cliff}}$
as $\mc K_{\text{cliff}}=\mc U_{\text{cliff}}\mc N$, where $\mc N$
represents the error channel that encompasses all errors. This decomposition
always exists when the dynamics of the gate constitute a completely
positive map. Substituting it into the twirling formula, we obtain:
\begin{align}
K_{\text{cliff}}^{RC} & =\frac{1}{2^{2n}}\sum_{\alpha=1}^{2^{2n}}\mc P_{\alpha}'\mc U_{\text{cliff}}\mc N\mc P_{\alpha}=\frac{1}{2^{2n}}\sum_{\alpha=1}^{2^{2n}}\mc P_{\alpha}'\mc U_{\text{cliff}}\mc P_{\alpha}\mc P_{\alpha}\mc N\mc P_{\alpha}\nonumber \\
 & =\frac{1}{2^{2n}}\sum_{\alpha=1}^{2^{2n}}\mc U_{\text{cliff}}\mc P_{\alpha}\mc N\mc P_{\alpha}=\mc U_{\text{cliff}}\frac{1}{2^{2n}}\sum_{\alpha=1}^{2^{2n}}\mc P_{\alpha}\mc N\mc P_{\alpha}.
\end{align}
Hence, twirling a noisy Clifford gate with $\mc P_{\alpha}$ and $\mc P_{\alpha}'$
is equivalent to twirling its noise channel $\mc N$ with the same
$\mc P_{\alpha}$ on both sides ($\mc P_{\alpha}'=\mc P_{\alpha}$).
Furthermore, we can utilize the fact that Pauli operators form a complete
basis, allowing to express any $\mc N$ as $\mc N=\sum_{\alpha,\beta}\mc N_{\alpha,\beta}P_{\alpha}\otimes P_{\beta}^{*}$
. It can be shown that, after Pauli twirling, $\mc N$ transforms
into 
\begin{equation}
\mc N^{RC}=\frac{1}{2^{2n}}\sum_{\alpha=1}^{2^{2n}}\mc P_{\alpha}\mc N\mc P_{\alpha}=\sum_{\alpha}\mc N_{\alpha,\alpha}\mc P_{\alpha}.
\end{equation}
In this transformation, only the diagonal elements remain. This resulting
channel is referred to as a Pauli error channel, and it corresponds
to randomly applying a Pauli unitary $\mc P_{\alpha}$ with a probability
of $\mc N_{\alpha,\alpha}$. Thus, regardless of the original structure
of $\mc N$, Pauli twirling turns it into a Pauli channel. As we show
later, this is not the case in PST.

\subsection{Introducing the Pseudo Twirling Protocol}

To introduce the concept of pseudo twirling we start with the following
example. Consider a simple non-Clifford gate, such as the two-qubit
gate $\mc U_{zz}(\theta)=e^{-i\theta\mc H_{zz}}$, with $\theta\neq k\pi/2$.
For simplicity, we use the notation $\mc H_{zz}=P_{zz}\otimes I-I\otimes P_{zz}^{\intercal}$
, $P_{zz}=\sigma_{z}\otimes\sigma_{z}$. In analogy to Pauli twirling
we aim to create an operation that keeps the functionality of the
ideal gate. If we apply $\mc P_{xx}=P_{xx}\otimes P_{xx}^{*}$, we
get: $\mc P_{xx}e^{-i\theta\mc H_{zz}}\mc P_{xx}=e^{-i\theta\mc H_{zz}}$
since $\mc P_{xx}$ and $\mc H_{zz}$ commute. However, $\mc P_{xz}$,
for example, anti-commutes with $\mc H_{zz}$, and therefore $\mc P_{xz}e^{-i\theta\mc H_{zz}}\mc P_{xz}=e^{+\imath\theta\mc H_{zz}}=\mc U_{zz}(-\theta)$.
Our PST protocol is based on the fact that if the Pauli operators
anti-commute with the Hamiltonian of $U$ (or equivalently $\mc U$
in Liouville space), we can simply change the sign of the angle, i.e.,
the sign of the driving fields, to achieve the desired transformation
in the end:

\begin{equation}
\mc P_{xz}\mc U_{zz}(-\theta)\mc P_{xz}=\mc U_{zz}(\theta).
\end{equation}

More generally when the driving Hamiltonian is not a single Pauli,
we can write $\mc U=e^{-i\sum h_{\beta}\mc H_{\beta}}$ and then the
PST twirling takes the form 
\begin{equation}
\mc K_{PST}=\frac{1}{2^{2n}}\sum_{\alpha=1}^{2^{2n}}\mc P_{\alpha}e^{-i\mc P_{\alpha}(\sum_{\beta}h_{\beta}\mc H_{\beta})\mc P_{\alpha}}\mc P_{\alpha}=\frac{1}{2^{2n}}\sum_{\alpha=1}^{2^{2n}}\mc P_{\alpha}e^{-i\sum_{\beta}sgn(\alpha,\beta)h_{\beta}\mc H_{\beta}}\mc P_{\alpha},\label{eq: PST}
\end{equation}
where $sgn(\alpha,\beta)=tr(P_{\alpha}P_{\beta}P_{\alpha}P_{\beta})/2^{n}$
equals $+1$ if $P_{\alpha}$ and $P_{\beta}$ commute, and $-1$
if they anti-commute. Note that in the definition of $sgn(\alpha,\beta)$
we used Pauli matrices in Hilbert space. Mathematically, expression
(\ref{eq: PST}) is straightforward, but it should be considered with
execution protocol in mind. The right $\mc P_{\alpha}$ is executed
at the beginning of the gate. Then a modified unitary $\mc U_{\alpha}=e^{-i\sum_{\beta}sgn(\alpha,\beta)h_{\beta}\mc H_{\beta}}$
is executed next by changing the sign of some of the control signals.
Finally, another $\mc P_{\alpha}$ is executed at the end of the gate.
Importantly, the Hamiltonian of the modified unitary has the same
elements as the original Hamiltonian, with only the sign of some terms
changing. Thus, under the assumption that the signs of the driving
Hamiltonian terms are controllable, the PST implementation is no more
challenging than implementing the original unitary.

The main operational differences between RC and pseudo twirling are:
\begin{itemize}
\item In RC, the Pauli gates at the beginning and at the end are different.
In pseudo twirling, they are always the same.
\item In RC, the operation between the Pauli gates (the original gate) is
always the same. Conversely, in PST, the operation between the Pauli
gates depends on the selected Pauli.
\item In both methods, the twirling gates are Pauli gates, but in PST, the
Pauli gates are also employed to determine the signs of the control
signals that drive the gate.
\end{itemize}
RC and PST are illustrated in Fig. \ref{fig: RC_PST}a and Fig. \ref{fig: RC_PST}b
respectively. We note that the term `pseudo twirling' reflects not
merely an operational difference from standard twirling but also its
distinct impact on errors, as discussed in the following two sections.
Its effects differs from Pauli twirling in two significant ways: i)
it cannot eliminate over/under rotations errors, and ii) in general,
it does not transform incoherent errors into a Pauli channel. Nevertheless,
we show and demonstrate that these differences do not diminish the
usefulness of pseudo twirling.

\subsection{PST Impact on Coherent Errors\label{subsec: PST-coh}}

In this section we study the impact of PST on coherent errors. Through
the Magnus expansion \citep{blanes2009magnus} in the interaction
picture, we find that the application of PST in a circuit with a coherent
error induced by a Hamiltonian term $\mc H_{coh}(t)$ leads to the
following evolution operator
\begin{equation}
\mc K_{PST}=\mc U_{T}\frac{1}{2^{2n}}\sum_{\alpha=1}^{2^{2n}}e^{-\imath\intop_{0}^{T}\mc U(t)^{\dagger}\mc P_{\alpha}\mathcal{H}_{coh}(t)\mc P_{\alpha}\mc U(t)dt},\label{eq: KpstExp}
\end{equation}
where $\mc U(t)$ is the ideal evolution operator of the non-twirled
gate and $\mc U_{T}=\mc U(T)$. The second and higher order terms
in the Magnus expansion have been neglected. $\mathcal{H}_{coh}$
is in Liouville space which is related to the Hilbert space Hamiltonian
$H_{coh}$ through Eq. (\ref{eq: Hlio}). The first order Taylor expansion
in $\mathcal{H}_{coh}$ is

\begin{align}
\mc U_{T}+\mc U_{T}\frac{-i}{2^{2n}}\sum_{\alpha=1}^{2^{2n}}\intop_{0}^{T}\mc U(t)^{\dagger}\mc P_{\alpha}\mathcal{H}_{coh}\mc P_{\alpha}\mc U(t)dt+O(\mathcal{H}_{coh}^{2}) & =\nonumber \\
\mc U_{T}+\mc U_{T}\frac{-i}{2^{2n}}\intop_{0}^{T}\mc U(t)^{\dagger}\sum_{\alpha=1}^{2^{2n}}\mc P_{\alpha}\mathcal{H}_{coh}\mc P_{\alpha}\mc U(t)dt+O(\mathcal{H}_{coh}^{2}) & =\mc U_{T}+O(\mathcal{H}_{coh}^{2}),\label{eq: PST1stCoh}
\end{align}
where the last step follows from the fact that $\frac{1}{2^{2n}}\sum_{\alpha=1}^{2^{2n}}\mc P_{\alpha}\mathcal{H}_{coh}\mc P_{\alpha}$
is a Pauli twirling of $\mathcal{H}_{coh}$. As such, the twirling
eliminates the leading term in the coherent error since $\sum_{\alpha=1}^{2^{2n}}P_{\alpha}H_{coh}P_{\alpha}=tr(H_{coh})I$
and $\sum_{\alpha=1}^{2^{2n}}P_{\alpha}^{*}H_{coh}^{\intercal}P_{\alpha}^{*}=tr(H_{coh})^{*}I=tr(H_{coh})I$
so that the two terms in $\mathcal{H}_{coh}$ cancel each other. This
explains why PST suppresses coherent errors. 

Next, we study the effect of PST on the second order Taylor approximation
in $\mathcal{H}_{coh}$. This term is important in light of the fact
the first order is eliminated by PST. Furthermore, this analysis will
unravel the structure of the resulting noise Channel. The second order
Taylor expansion of (\ref{eq: KpstExp}) leads to following 2nd order
term in $\mc K_{pst}$
\begin{equation}
\mc K_{PST}=\mc U_{T}+\mc U_{T}\frac{1}{2^{2n}}\sum_{\alpha=1}^{2^{2n}}\frac{1}{2}[-i\intop_{0}^{T}\mc U(t)^{\dagger}\mc P_{\alpha}\mathcal{H}_{coh}\mc P_{\alpha}\mc U(t)dt]^{2}+O(\mathcal{H}_{coh}^{3}).
\end{equation}

We start by observing that $\intop_{0}^{T}\mc U(t)^{\dagger}\mc P_{\alpha}\mathcal{H}_{coh}\mc P_{\alpha}\mc U(t)dt$
still has the structure of a Hamiltonian in Liouville space (\ref{eq: Hlio})
$H_{eff,\alpha}\otimes I-I\otimes H_{eff,\alpha}^{\intercal}$ where
$H_{eff,\alpha}=\intop_{0}^{T}U(t)^{\dagger}P_{\alpha}H_{coh}P_{\alpha}U(t)dt$.
Using this observation the expression $[-i\intop_{0}^{T}\mc U(t)^{\dagger}\mc P_{\alpha}\mathcal{H}_{coh}\mc P_{\alpha}\mc U(t)dt]^{2}$
now reads
\begin{align}
-(H_{eff,\alpha}\otimes I-I\otimes H_{eff,\alpha}^{\intercal})^{2} & =2H_{eff,\alpha}\otimes H_{eff,\alpha}^{\intercal}-H_{eff,\alpha}^{2}\otimes I-I\otimes(H_{eff,\alpha}^{2})^{\intercal}.
\end{align}
Since $H_{eff,\alpha}$ is Hermitian this expression has a Lindblad
form $\mc L(A)=A\otimes A^{*}-\frac{1}{2}(A^{\dagger}A)\otimes I-\frac{1}{2}I\otimes(A^{\dagger}A)^{\intercal}$
with a Hermitian dissipator $A=H_{eff,\alpha}$. We conclude that
PST nulls the first order of the coherent error and the second order
produces a Hermitian Lindbladian. That is 
\begin{align}
\mc K_{PST} & \simeq\mc U_{T}e^{\mc L_{eff}},\\
\mc L_{eff} & =\frac{1}{2^{2n}}\sum_{\alpha=1}^{2^{2n}}\mc L(H_{eff,\alpha}).\label{eq: LeffCoh}
\end{align}
Since $\mc L_{eff}=\mc L_{eff}^{\dagger}$ the resulting noise channel
$\mc N_{PST}=\mc U_{T}^{\dagger}\mc K_{PST}\simeq e^{\mc L_{eff}}$
is also Hermitian up to higher order corrections. We emphasize that
unlike RC, $\mc N_{PST}$ is not a Pauli channel although it is approximately
Hermitian. This is easily observed when plotting $\mc N_{PST}$ in
the Pauli basis. The noise map is not diagonal as in RC. This fundamental
difference highlights that PST is not a variant of RC or a specific
way of implementing RC. It is a different protocol with different
analytical features and a different operating regime.

If $H_{coh}$ is a Pauli then $\mc P_{\alpha}\mathcal{H}_{coh}\mc P_{\alpha}=\pm\mathcal{H}_{coh}$.
However since $\mc L(A)=\mc L(-A)$ we get that $\mc L_{eff}$ simplifies
to 
\begin{equation}
\mc L_{eff}=\mc L[\intop_{0}^{T}U(t)^{\dagger}H_{coh}U(t)dt].
\end{equation}
We leave the study of higher order terms to future research. 

\subsection{\label{subsec: PST-and-Noise}PST and Noise}

In the context of Pauli twirling, it is evident that the resulting
incoherent error exhibits a Pauli error channel structure. However,
in the case of PST, this may not hold true, even in the first order.
If we repeat the analysis from the previous section but consider a
general Lindblad dissipator $\mathcal{L}$ describing a Markovian
trace-preserving noise, we obtain
\begin{equation}
\mc U_{T}+\mc U_{T}\frac{-i}{2^{2n}}\intop\mc U(t)^{\dagger}\sum_{\alpha=1}^{2^{2n}}\mc P_{\alpha}\mathcal{L}\mc P_{\alpha}\mc U(t)dt+O(\mathcal{L}^{2}),
\end{equation}
where $\mathcal{L}=\sum_{k}A_{k}\otimes A_{k}^{*}-\frac{1}{2}A_{k}^{\dagger}A_{k}\otimes I-\frac{1}{2}I\otimes(A_{k}^{\dagger}A_{k})^{\intercal}$
is the native noise. The Pauli twirling of $\mathcal{L}$ turns it
into a Pauli channel, however, the summation over additional rotation
by $\mc U(t)^{\dagger}$ make the whole expression a non Pauli channel.
Yet, since $\mathcal{L}'=\sum_{i=1}^{2^{2n}}\mc P_{i}\mathcal{L}\mc P_{i}$
is Pauli channel, it is Hermitian, and consequently, $\intop\mc U(t)^{\dagger}\mathcal{L}'\mc U(t)dt$
is Hermitian as well. We conclude that PST makes the leading order
of the noise Hermitian. Please refer to our numerical results section
for supporting simulations. We observe that the anti-Hermitian part
of the noise, can decrease by a factor of several hundreds or more
after applying PST. The impact of PST on higher-order noise terms
is left for future work. Note that PST does not only mitigate the
non-Hermitian part of the noise it also modifies the original Hermitian
component of the noise.

Interestingly, in \citep{HermitanNoisePhseEst2023} it was shown that
to a leading order, Hermitian noise does not degrade the result of
control-free phase estimation. To exploit this feature, the authors
applied randomized compiling for transforming the noise into a Pauli
channel (which is Hermitian). However this technique can only be applied
to Clifford gates. PST can pose an alternative for making the leading
order of the noise Hermitian even for short low-noise non-Clifford
gates.

\section{PST features - limitations and opportunities}

\subsection{\label{subsec: PST-and-Calibration}PST, Additional Types of Coherent
Errors, and Calibration }

Coherent errors comprise of either controllable errors arising from
miscalibration of controllable degrees of freedom, or uncontrollable
errors such as crosstalk that do not depend on the control signals
(driving fields). While RC makes no distinction and effectively mitigate
both types of coherent errors, PST addresses all coherent error excluding
controllable rotation errors (over or under rotation) which play a
key role in gate calibration. This feature opens new possibilities. 

At first glance, it may seem that this limitation renders PST only
a partial solution, as it cannot transform controllable rotation errors
into incoherent errors. However, this exact limitation can become
an advantage in gate calibration. A proper calibration is always a
good practice. When RC is applicable, it will convert the miscalibration
error into excessive noise that makes the QEM task more challenging.
PST, on the other hand, does not convert controllable rotation errors
into incoherent errors. This characteristic allows for the application
of PST during the calibration process. By reducing all other coherent
errors, and leaving only those associated with the calibration itself,
it simplifies the calibration process and improves its accuracy. The
high accuracy calibration that PST facilitates, is imperative for
circuits with significant depth. 

As a concrete scenario we consider a simplified cross-resonance Hamiltonian
$H_{d}=a\cos\phi\,\sigma_{z}\otimes\sigma_{x}+a\sin\phi\,\sigma_{z}\otimes\sigma_{y}+\zeta\sigma_{z}\otimes\sigma_{z}$,
where $a$ is the amplitude of the drive, $\phi$ is the phase of
the drive, and $\zeta$ is the crosstalk coupling between the qubits.
See \citep{McKa2020CRgateFirstP} for additional terms. Since $\sigma_{z}\otimes\sigma_{x}$
is the target Hamiltonian, a deviation of the coefficient $a\cos\phi$
from the desired value is considered as a rotation error. Since it
is possible to change the sign of this coefficient we consider it
as a controllable rotation error. The term $a\sin\phi\,\sigma_{z}\otimes\sigma_{y}$
is a controllable non-rotational error and the last crosstalk term
is an uncontrollable non-rotational error. Such scenarios motivate
a more formal and accurate classification of coherent errors. Let
us assume we have a Hamiltonian of the form
\begin{equation}
\mc H=f(a)\mc H_{\beta}+\sum_{\gamma\neq\beta}g_{\gamma}(a)\mc H_{\gamma},
\end{equation}
where $\mc H_{\beta}$ is the target Pauli Hamiltonian (multi-qubit
Pauli), $\mc H_{\gamma}$ represents Pauli coherent error terms, and
$a$ is the parameter being controlled. Without loss of generality
we assume $\mc H_{\gamma}\neq\mc H_{\beta}$. If they are equal it
is possible to absorb $g(a)$ in the definition of $f(a)$. For brevity,
in what follows we consider a single $\mc H_{\gamma}$ term. Next,
we decompose $f$ and $g$ into even and odd contributions, i.e. $f(a)=f_{e}(a)+f_{o}(a)$
where $f_{e}(a)=\frac{1}{2}[f(a)+f(-a)]$ and $f_{o}(a)=\frac{1}{2}[f(a)-f(-a)]$.
We refer to the odd terms as controllable since changing the sign
of $a$ changes the sign of the terms in the Hamiltonian as needed
for PST. Similarly, the even functions represent uncontrollable degrees
of freedom from the point of view of PST. Miscalibration of $f(a)$,
regardless if the odd or even term is the problem, leads to rotation
errors. $g(a)$ stands for non-rotation errors that stems, for example,
from crosstalk or parasitic terms in the control signals that generate
the target term $\mc H_{\beta}$. In what follows, it is shown that
PST mitigates the terms proportional to $f_{e}(a),g_{o}(a),$ and
$g_{e}(a)$, while $f_{o}(a)$ remains unaffected. 

While $f_{e}(a),g_{o}(a),$ and $g_{e}(a)$ are small since they represent
coherent errors, $f_{o}(a)$ may be potentially large. This is addressed
by moving to the interaction picture of $f_{o}(a)\mc H_{\beta}$.
After applying PST using $\mc{P_{\alpha}}$ and $a\to sgn(\alpha,\beta)a$,
the first order expansion {[}as in (\ref{eq: PST1stCoh}){]}, yields

\begin{align*}
\mc K_{PST} & =\mc U_{T}^{o}+\mc U_{T}^{o}\frac{-i}{2^{2n}}\intop_{0}^{T}\mc U^{o}(t)^{\dagger}\sum_{\alpha=1}^{2^{2n}}\mc P_{\alpha}\{f_{e}(a)\mathcal{H}_{\beta}+[g_{e}(a)+sgn(\alpha,\beta)g_{o}(a)]\mathcal{H}_{\gamma}\}\mc P_{\alpha}\mc U^{o}(t)dt\\
 & +O(f_{e}^{2},g_{e}^{2},g_{o}^{2},g_{e}g_{o},f_{e}g_{o},f_{e}g_{e}),
\end{align*}
where we have used the parities of $f_{e}$, $g_{e}$ and $g_{o}$
to simplify the expression. $\mc U^{o}(t)$ is the evolution operator
generated by $f_{o}(a)\mc H_{\beta}$ ($\mc U_{T}^{o}=\mc U^{o}(T))$.
$\mc U_{T}^{o}$ contains all potential controlled rotation errors.
We can now focus on the effective Hamiltonian:
\begin{align}
\mc H_{eff} & =\frac{1}{2^{2n}}\sum_{\alpha=1}^{2^{2n}}\mc P_{\alpha}\{f_{e}(a)\mathcal{H}_{\beta}+[g_{e}(a)+sgn(\alpha,\beta)g_{o}(a)]\mathcal{H}_{\gamma}\}\mc P_{\alpha}\\
 & =\frac{g_{o}(a)}{2^{2n}}\sum_{\alpha=1}^{2^{2n}}sgn(\alpha,\beta)\mc P_{\alpha}\mc H_{\gamma}\mc P_{\alpha},
\end{align}
where we have used the fact that $f_{e}(a)$ and $g_{e}(a)$ do not
depend on $\alpha$ and therefore the regular twirling property $\sum_{\alpha=1}^{2^{2n}}\mc P_{\alpha}\mc H_{\beta(\gamma)}\mc P_{\alpha}=0$
nulls these terms. In Appendix II, we show that when $\mc H_{\gamma}\neq\mc H_{\beta}$
it holds that $\sum_{\alpha=1}^{2^{2n}}sgn(\alpha,\beta)\mc P_{\alpha}\mc H_{\gamma}\mc P_{\alpha}=0$
and we get $\mc H_{eff}=0$ so that

\begin{equation}
\mc K_{PST}=\mc U_{T}^{o}+O(f_{e}^{2},g_{e}^{2},g_{o}^{2},g_{e}g_{o},f_{e}g_{o},f_{e}g_{e}).
\end{equation}
In the cross-resonance case discussed above $f_{o,zx}(a)=a\cos\phi$,
$g_{o,zx}(a)=a\sin\phi$, $f_{e,zx}=0,g_{e,zx}=0$,$g_{o,zz}=0$ and
$g_{e,zz}=\zeta$. After PST the effective driving Hamiltonian is 

\begin{align}
\mc H_{d}^{PST} & =a\cos\phi\mc H_{zx}.
\end{align}
This leads to a simplification in the calibration procedure: rather
than fine-tune $\phi$ to zero and then fine-tune $a$ (e.g. to $\pi/4$),
it is possible to coarsely set $\phi\sim0$ and then calibrate $a$
until $a\cos\phi$ achieves the desired value. Furthermore, PST also
suppresses crosstalk to neighboring qubits which improves the accuracy
of calibration.

\subsection{Alternative Ways for Applying PST in a Circuit}

Since PST is versatile and can be applied to any gate or circuit,
there are various alternatives for its application within a given
circuit. Some options include:
\begin{itemize}
\item Edges PST - Applying PST to the entire circuit, with the twirling
gates positioned solely at the circuit's edges.
\item Gate PST- Applying PST on a per-gate basis.
\item Intra-gate PST - Implementing PST on elements within each gate.
\item Hybrid PST-RC - In Clifford gates, combine Intra-gate PST with RC,
or RC with edges PST.
\end{itemize}
The first three options are illustrated in Fig. \ref{fig: RC_PST}c.
In principle, it appears advantageous to consider all three levels
of implementation, or at the very least, the first and third options.
Implementing PST in shorter intervals (intra-gate PST) is an effective
strategy to prevent the accumulation of coherent errors. Employing
PST across the entire circuit is also beneficial, as it can mitigate
the accumulation of errors resulting from non-local (non-gate-based)
mechanisms.\\
\textbf{Intra-Gate PST}

\begin{figure}
\begin{centering}
\includegraphics[width=12cm]{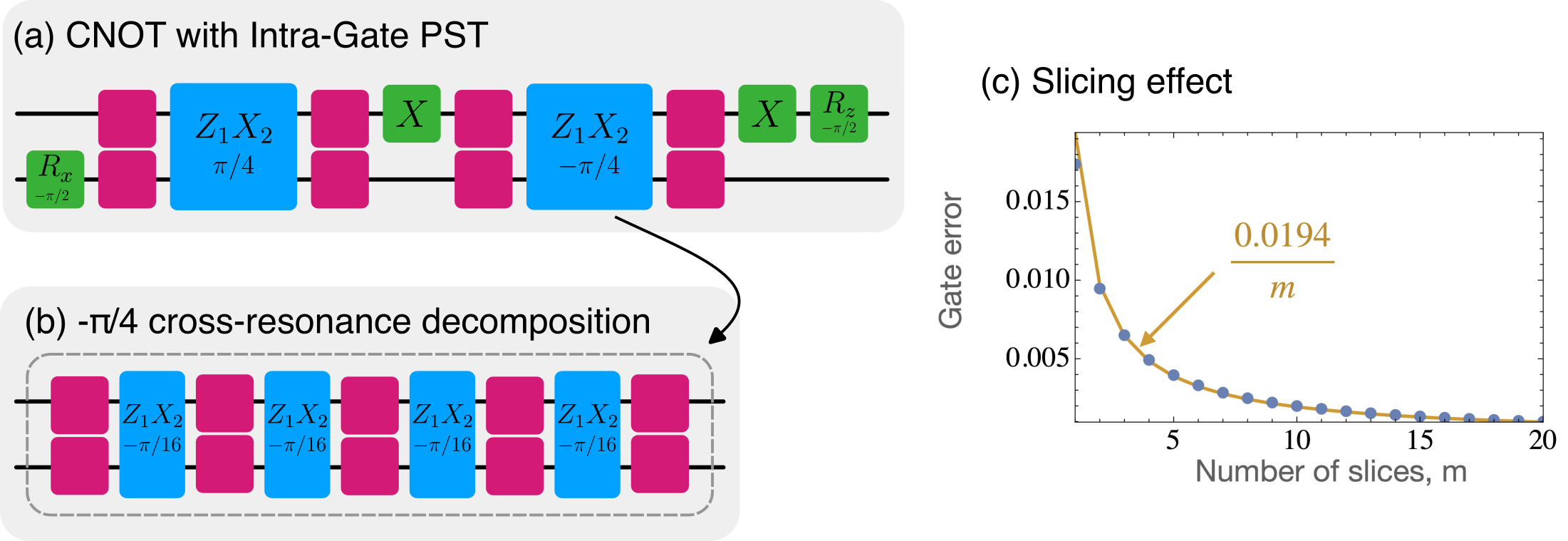}
\par\end{centering}
\caption{\label{fig: intra gate}(a) A schematic representation of the pulse
schedule that generates the IBM echo cross-resonance gate. (b) Intra-gate
PST: since PST is not limited to Clifford gates, it can be applied
separately to each of the ZX pulses (red lines). (c) Taking it a step
further, it is possible to decompose each of the \textgreek{p}/4 ZX
pulses into $m$ shorter pulses and apply PST to each of these shorter
pulses. (d) The slicing effect on a $\pi/2$ ZX pulse. The x-axis
represents the number of slices (pulses), while the y-axis indicates
the operator norm error from the ideal pulse $\protect\nrm{\protect\mc K_{PST}^{(m)}-\protect\mc U_{T}}_{op}$.
The solid orange line corresponds to a fit using the function $b/m$,
where $b$ is a fitting parameter.}

\end{figure}

Let us elaborate on the intra-gate PST approach. Taking the example
of the cross-resonance CNOT gate \citep{McKa2020CRgateFirstP} depicted
in Fig. \ref{fig: intra gate}a, each non-Clifford element (pulse)
can be pseudo twirled independently within the PST framework as shown
in Fig. \ref{fig: intra gate}b. This contrasts with RC, which cannot
be applied to the non-Clifford elements of the gate. Additionally,
each one of two segments of the cross-resonance pulse can be divided
into multiple shorter pulses, allowing for the application of PST
across these smaller segments (Fig. \ref{fig: intra gate}c). The
prospective advantage here is to mitigate coherent errors before they
accumulate. To quantify this claim we show in Appendix III that when
the number of slices $m$ is sufficiently large so that neither the
drive nor the coherent error changes on the scale of a single slice,
the evolution operator is given by:
\begin{equation}
\mc K_{PST}^{(m)}=\mc U_{T}e^{\frac{T}{m}\intop\mc U^{\dagger}(t)\frac{1}{2^{2n}}\sum_{\alpha=1}^{2^{2n}}\mc L(P_{\alpha}H_{coh}P_{\alpha})\mc U(t)dt}.\label{eq: K(m)}
\end{equation}
While the integral is $m$ independent, the coefficient before it
scales like $1/m$ which agrees with the numerical observations in
Sec. \ref{subsec: Intra-Gate-PST}. We can quantify the residual noise
using the operator norm (see Appendix IV for the explanation why this
norm is used)
\begin{equation}
\nrm{\mc K_{PST}^{(m)}-\mc U_{T}}_{op}\simeq\frac{T}{m}\nrm{\intop\mc U^{\dagger}(t)\frac{1}{2^{2n}}\sum_{\alpha=1}^{2^{2n}}\mc L(\mc{P_{\alpha}}\mc H_{coh}\mc{P_{\alpha}})\mc U(t)dt}_{op},\label{eq: Km err norm}
\end{equation}
which clearly scales like $1/m$. 

In the context of Clifford gates such as the CNOT, conventional RC
can be applied at the gate level, while employing PST within the gate's
internal elements. As PST mitigates the non-rotational coherent errors,
RC will have to cope mostly with the residual rotation errors.

\subsection{Integration and Compatibility with Quantum Error Mitigation }

It is evident that PST, by itself, will only provide a limited solution
to errors in quantum devices, as it converts coherent errors into
incoherent errors, which still need to be addressed using QEM. Therefore,
the impact of PST largely depends on its compatibility with QEM methods.
PST was mainly introduced for handling non-Clifford gates. Yet QEM
methods that are restricted to Cliffords gates such PEC and PEA may
still benefit from intra-gate PST and circuit level PST. 

Finally, we point out that much like the `Adaptive KIK' QEM method
\citep{KIKarxiv}, the feasibility of executing PST hinges on the
ability to control the signs of the different terms in the driving
Hamiltonian. We have shown in Section \ref{subsec: PST-coh} and \ref{subsec: PST-and-Noise}
that PST turns the non-rotation coherent errors and non Hermitian
dissipators into a Hermitian Lindblad dissipator. Hence it is expected
that PST will be compatible with the Adaptive KIK method. In Sec.
\ref{subsec: Integration-of-PST-QEM} and \ref{sec: Experimental-Results},
we demonstrate numerically and experimentally that PST is not only
compatible with Adaptive KIK, but substantially improves the resilience
of the method to crosstalk. 

\section{Numerical Results}

\subsection{Intra-Gate PST\label{subsec: Intra-Gate-PST}}

As alluded to in Fig. \ref{fig: intra gate}c, implementing PST on
smaller segments of a pulse allows for further improvement in coherent
error suppression. To illustrate this point, we conducted a simulation
involving the slicing of a $\pi/2$ cross-resonance pulse. The uncontrolled
coherent error is a $\sigma_{z}\otimes\sigma_{z}$ term that accompanies
the $\sigma_{z}\otimes\sigma_{x}$ driving Hamiltonian. We have applied
PST to each one of the $m$ slices and plotted the error in the resulting
evolution operator $\mc K_{PST}^{(m)}$ with respect to the ideal
evolution $\mc U_{ideal}$ of a $\pi/2$ cross-resonance pulse. Figure
\ref{fig: intra gate}d shows that the resulting error, which we quantify
using $\nrm{\mc K_{PST}^{(m)}-\mc U_{ideal}}_{op}$, reduces as the
number of slices increases. In agreement with (\ref{eq: K(m)}) we
observe that the level of suppression is inversely proportional to
the number of slices. The orange curve is a fit of the function $b/m$
to the data, where $b$ is a fit parameter that corresponds to the
norm operator of the first Magnus term of the noise {[}see (\ref{eq: Km err norm}){]}.
Intuitively, in thin slices the PST mitigates the coherent error before
it has a chance to accumulate. 

In practice, at some point the duration of the slice becomes comparable
to that of the PST Pauli operators, and the small error in these operators
sets a performance limit. Nonetheless, our results suggest that before
reaching this limiting factor intra-gate PST improves the coherent
error suppression.

\subsection{PST as a ``Hermitianizer'' of the Error Channel}

\begin{figure}[h]
\begin{centering}
\includegraphics[width=7cm]{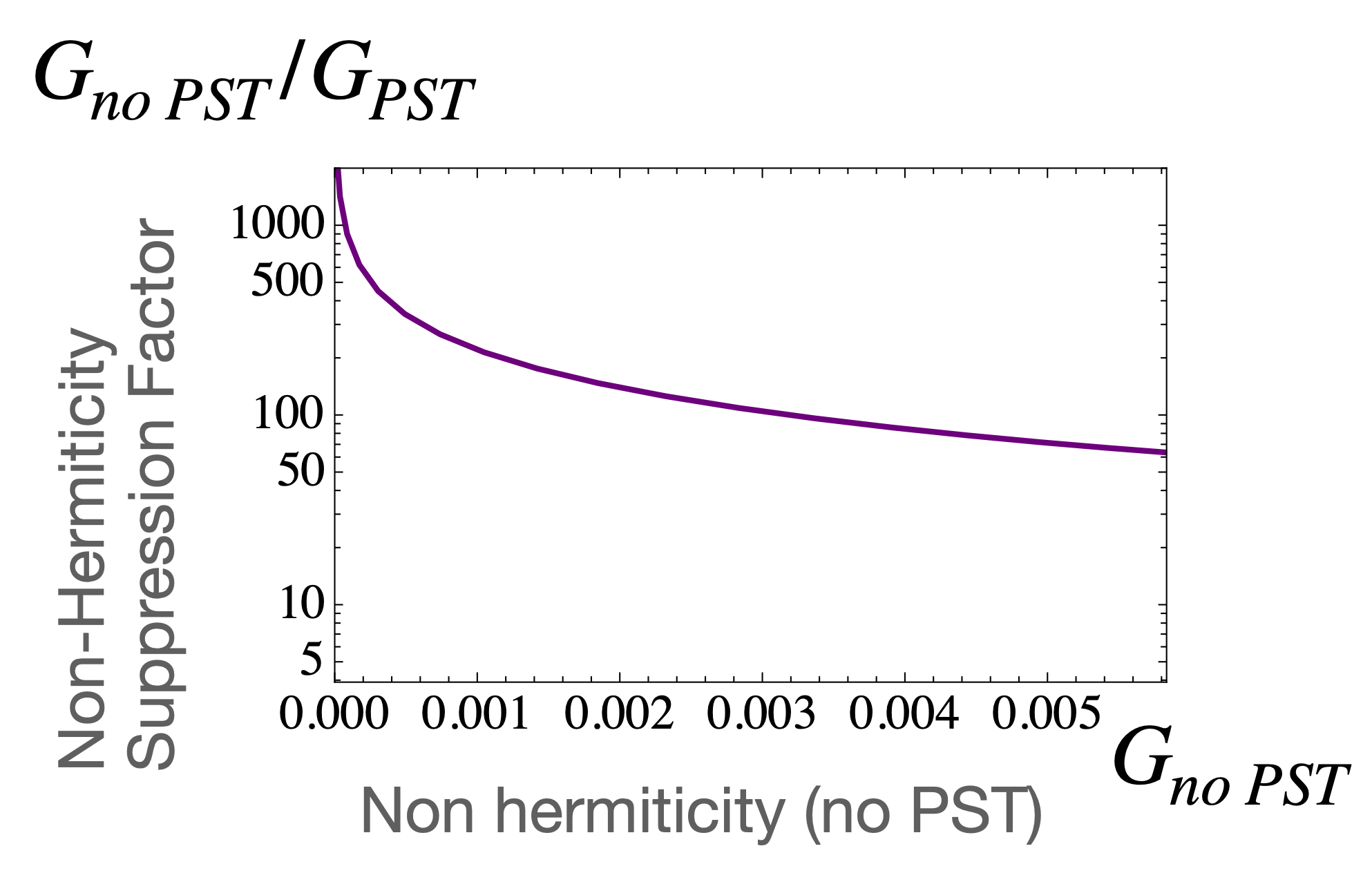}\caption{\label{fig: hermitianizer}The effect of PST on incoherent errors
(noise). The setup is the same as the cross-resonance pulse setup
presented in Fig. \ref{fig: intra gate}, but this time with incoherent
errors instead coherent errors. The $y$ axis shows the non-Hermiticity
suppression factor $G_{no\:PST}/G_{PST}$ where $G=\frac{1}{2}\left\Vert \protect\mc N-\protect\mc N^{\dagger}\right\Vert _{op}/\left\Vert \protect\mc N\right\Vert _{op}$
measure the non-Hermiticity of the noise channel $\protect\mc N$. }
\par\end{centering}
\end{figure}

To illustrate the effect of PST on incoherent errors, we consider
again a $\pi/2$ echo cross-resonance pulse, but this time with coherent
errors replaced by incoherent errors. A $T_{1}$ noise (amplitude
damping) is applied to each qubit. To quantify the non-Hermiticity
of a noise channel $\mc N=\mc U^{\dagger}\mc K$ we use the measure
$G=\left\Vert \mc N-\mc N^{\dagger}\right\Vert _{op}/\left\Vert 2\mc N\right\Vert _{op}$.
$G$ equals one for anti-Hermitian operators, and zero for Hermitian
operators. Figure \ref{fig: hermitianizer} displays how the PST suppresses
$G$ for different level of initial non-Hermiticity. The initial non-Hermiticity
was controlled by varying the strength of the amplitude damping noise.
As a result, the Hermitian part of the noise increases as well. We
have observed the same behavior when using randomly chosen Lindbladians
from an ensemble of random Lindbladians \citep{randLind}.

\begin{figure}[h]
\begin{centering}
\includegraphics[width=15cm]{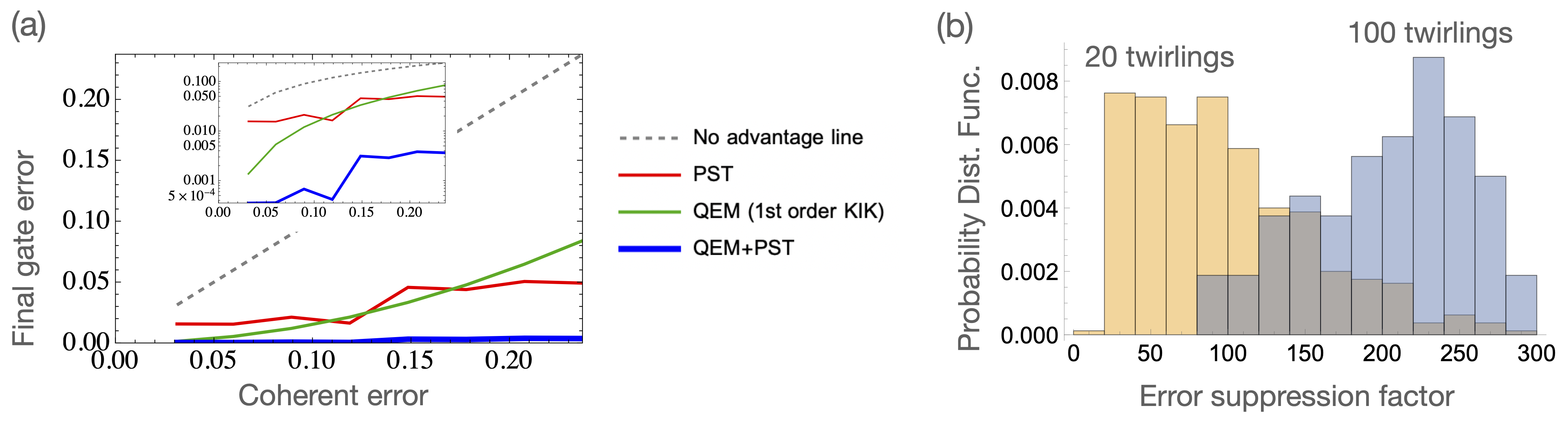}
\par\end{centering}
\caption{\label{fig:PSTKIK}(a) Integration of pseudo twirling with the Adaptive
KIK quantum error mitigation method. Different $x$ values represent
the operator norm error of the evolution operator when implementing
a 3-qubit transverse Ising time-evolution. The error increase in the
$x$ axis is obtained by increasing the coherent error amplitude.
The $y$-axis shows the operator norm error after the application
of various protocols. The dashed line represents errors without any
technique applied. This is a `break-even' line where any result below
it represents an improvement. The green curve corresponds to the first-order
Adaptive KIK mitigation without PST and the red curve depicts the
error reduction when PST is applied without KIK. The combination of
KIK and PST (blue curve) showcases a significant error suppression.
The inset displays the same plot in a logarithmic scale. For each
data point we have used twenty twirlings. (b) Using a hundred twirlings,
for initial coherent error of 0.25 (corresponding to the right edge
of Fig. (a)) substantially improves the error suppression of KIK+PST
as shown by the blue histogram.}
\end{figure}

\subsection{\label{subsec: Integration-of-PST-QEM}PST+QEM Integration in the
Ising Model}

As mentioned earlier, PST alone provides only a partial solution to
errors in quantum computers. Ideally, it transforms the noise and
coherent errors into a Hermitian noise channel. However, the resulting
noise channel can still be too strong to extract meaningful results
from a given quantum computer. Therefore, PST must be complemented
by a quantum error mitigation protocol. We choose the `Adaptive KIK'
QEM method \citep{KIKarxiv} due to its operational simplicity and
its compatibility with non-Clifford gates. To exemplify the compatibility
of PST with this QEM method, we conducted a three-qubit transverse
Ising model simulation with both coherent and incoherent errors. The
coherent error Hamiltonian was chosen to be $\{\sigma_{z,i}\}$. Such
an error can arise from imprecise determination of the qubits frequencies.
Finally, the evolution before the PST is given by:
\begin{equation}
\mc K=e^{-i\,J\sum_{i=1}^{2}\mc H_{z_{i},z_{i+1}}-i\,g\sum_{i=1}^{3}\mc H_{x_{i}}-i\,\epsilon\sum_{i=1}^{3}\mc H_{z_{i}}+\mc L},
\end{equation}
where $J$ and $g$ are parameters of the transverse Ising Hamiltonian
and $\epsilon$ sets the amplitude of the coherent error. The pulse
inverse $\mc K_{I}$ needed for the KIK protocols is 

\begin{equation}
\mc K_{I}=e^{+i\,J\sum_{i=1}^{2}\mc H_{z_{i},z_{i+1}}+i\,g\sum_{i=1}^{3}\mc H_{x_{i}}-i\,\epsilon\sum_{i=1}^{3}\mc H_{z_{i}}+\mc L}.
\end{equation}
Importantly, in the pulse inverse $\mc K_{I}$ the sign of $\epsilon$
remains the same since it represents an uncontrollable coherent error
that does not depend on the external drive. The first order KIK mitigated
operator is 
\begin{equation}
\mc K^{(1)}=\frac{3}{2}\mc K^{PST}-\frac{1}{2}\mc K^{PST}\mc K_{I}^{PST}\mc K^{PST}.
\end{equation}
Figure \ref{fig:PSTKIK}(a) illustrates that PST significantly enhances
the KIK performance, as evident from the shift from the green curve
to the blue curve. To emphasize the insufficiency of PST alone, we
have also plotted in red the error when PST is applied without KIK.
At the rightmost point the PST+KIK reduces the error by a factor of
65.

In this simulation we randomly sampled the twirling space and used
only 20 twirling instances. In the implementation of the evolution
$\mc K$ and $\mc K\mc K_{I}\mc K$ needed for the KIK protocol we
used the same twirlings for all $\mc K$ and $\mc K_{I}$. Figure
\ref{fig:PSTKIK}(b) shows the dependence of the error suppression
ratio of the PST combined with 1st order KIK protocol with respect
to the original error for $n_{PST}=20$ twirlings (orange) and $n_{PST}=100$
twirlings (blue) . The original noise level is 0.25 (right hand side
of Fig. \ref{fig:PSTKIK}(a)). These histograms were generated by
repeating the simulation 800 times with different random twirlings.
For the simulation parameter we used $J=0.1,g=0.2,\xi=0.00465$. $\epsilon$
was varied from $0.005$ to $0.05$. For the dissipative term $\mc L$
we use $T_{1}$ noise (amplitude damping) for the different qubits
such that $\mc L=0.5\mc L_{1}^{T_{1}}+1.7\mc L_{2}^{T_{1}}+0.3\mc L_{3}^{T_{1}}$
where the subscript designates the qubit index. The initial state
is $\ket{000}$. The the error was quantified using the operator norm
as in Fig. 5.

\section{Experimental Results\label{sec: Experimental-Results}}

To illustrate the utility of PST to gate-calibration we run a `deep
calibration' experiment where we calibrate a sequence of 81 echo cross-resonance
(ZX) $\pi/2$ pulses. Such a large number of repetitions amplifies
the smallest coherent errors. Yet, a trivial execution of such a long
circuit is prone to substantial incoherent errors (noise) that will
dramatically effect the calibration curve and lead to an erroneous
calibration as demonstrated in \citep{KIKarxiv}. Thus, we use the
Adaptive KIK \citep{KIKarxiv} QEM methodology for removing the incoherent
errors. We measure the probability to find the system in the initial
$\ket{00}$ state (survival probability, SP) after the 81 repetitions
sequence for different stretch factor $\chi$ of the cross-resonance
pulse amplitude. $\chi=1$ corresponds to the default amplitude value
provided by IBM. In this experiment we use the \emph{ibmq\_kolkata}
quantum computer.

\begin{figure}
\begin{centering}
\includegraphics[width=14cm]{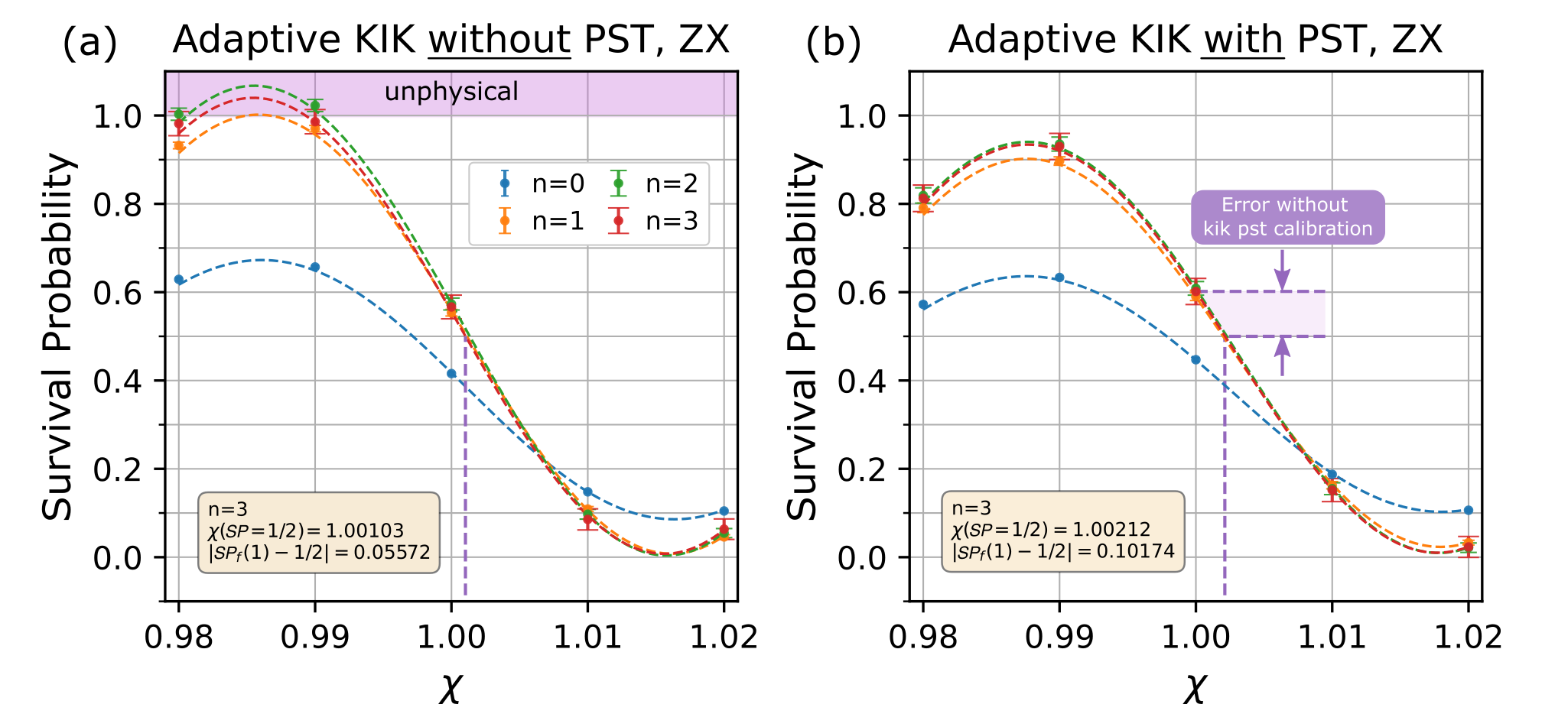}
\par\end{centering}
\caption{\label{fig: PST_calib_ZX}A `deep calibration' experiment of a cross-resonance
gate is conducted on \emph{ibmq\_kolkata}. The initial state is set
to $\protect\ket{00}$, then a sequence of 81 $\pi/2$ echo cross-resonance
pulses is executed, aiming at amplifying small coherent errors. The
$x$ axis describes the cross-resonance amplitude stretch factor $\chi$.
At such depths, the incoherent errors become pronounced, and error
mitigation has to be applied. $n=0$ corresponds to no error mitigation,
and $n>0$ corresponds to higher-order Adaptive KIK error mitigation
methods. The point $\chi(SP=0.5)$, where the graph intersects the
value $SP=0.5$ (vertical lines), sets the stretch factor that should
be used to recalibrate the amplitude of the cross-resonance pulse.
The left figure (a) shows the results without PST, and the right figure
(b) shows the results with PST using twenty random twirlings. Not
only does PST speeds up the convergence ($n=2$ suffices with PST),
but PST also prevents the unphysical results observed without PST
(shaded-purple region, probability larger than one). These non-physical
results can arise due to non-Markovian crosstalk effects, which PST
suppresses. Although the vertical lines indicate that the amplitude
should be modified only by $0.2\%$ with respect to the default value
set by IBM ($\chi=1$), it is observed that the error in the survival
probability is as big as $0.1$ if $\chi=1$ is used instead of $\chi(SP=1/2)\simeq1.0021$.}

\end{figure}

By comparing the results of applying KIK with and without PST we observe
in Fig. \ref{fig: PST_calib_ZX} that i) with PST (Fig. \ref{fig: PST_calib_ZX}b)
the KIK converges faster (less mitigation orders are needed) compared
to the same experiment without PST (Fig. \ref{fig: PST_calib_ZX}a).
ii) The value of the calibrated amplitude $\chi(SP=1/2)$ is affected
by the PST. iii) Without PST for some $\chi$ values we observe unphysical
values. To the best of our knowledge, such non-physical values can
occur for two reasons. First, if the mitigation order is too small
with respect to the window size parameter in the KIK method (see simulations
in the appendix of \citep{KIKarxiv}) nonphysical values can appear.
Yet, in this case, it is expected that the unphysical values will
disappear in higher mitigation orders. The other reason is pronounced
non-Markovian effects. One of the sources of non-Markovian effects
in the present setup is crosstalk interaction with adjacent qubits.
$n=3$ appears to be more physical than $n=2$, yet, since the two
results differ, convergence is not observed and we cannot confirm
that unphysical values disappear in higher orders. Moreover, when
the mitigation order is too low PST is not expected to help. The fact
that PST removes the unphysical values and speeds up convergence suggests
that the unphysical values without PST are related to non-Markovian
effects. We point out that in other circuits we have suppressed unphysical
values without PST by applying dynamical decoupling to neighboring
idle qubits.

While we control $\chi$, the actual drive of the ZX term is $\chi a\cos(\phi)$
where $\phi$ is the phase of the drive. In PST calibration it is
not needed to know $\phi$ explicitly, $\chi$ is scanned until the
survival probability obtain the value of $1/2$ which means that $\chi a\cos(\phi)=\pi/4$
(the value that corresponds to $\pi/2$ cross-resonance pulse). 

\begin{figure}
\begin{centering}
\includegraphics[width=14cm]{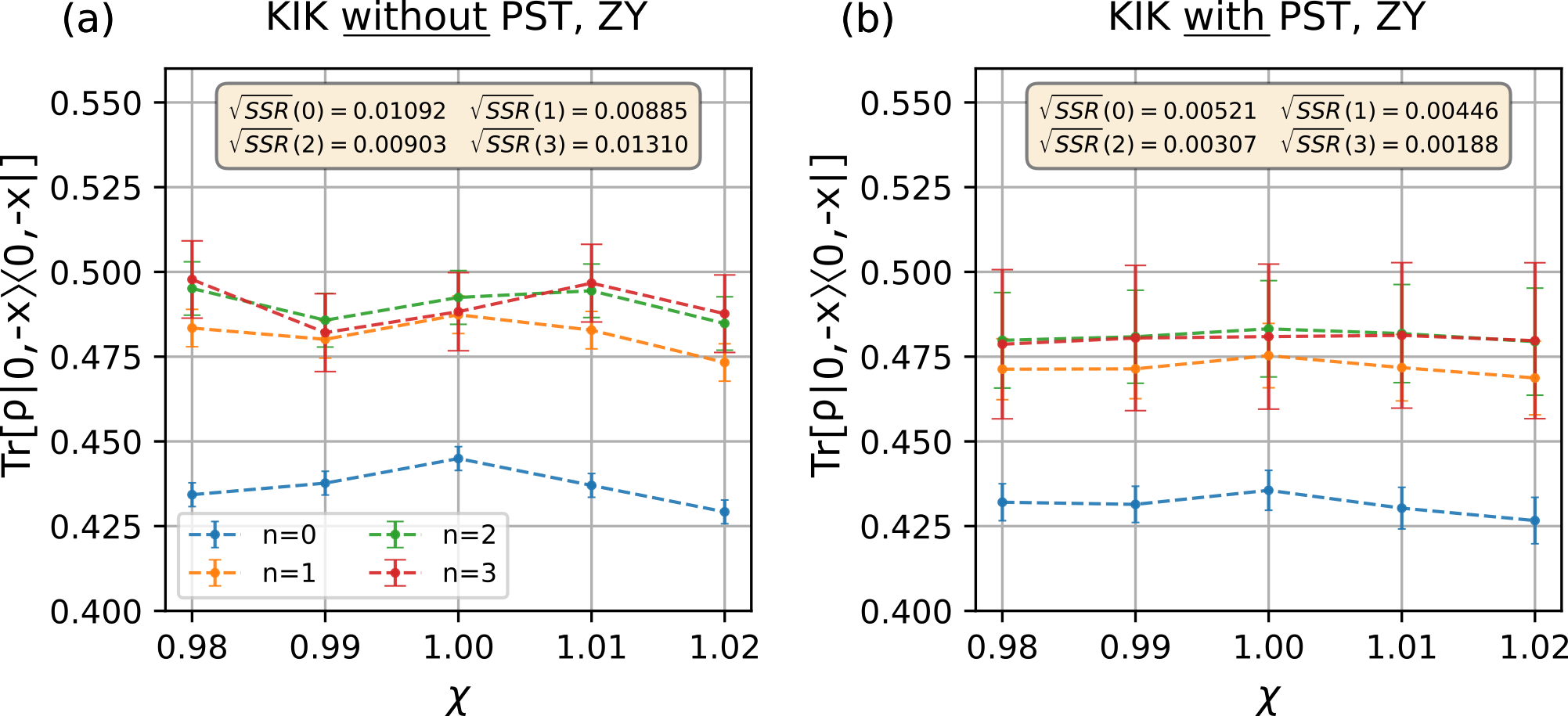}
\par\end{centering}
\caption{\label{fig: PST_calib_ZY}The same experiment as shown in Fig. \ref{fig: PST_calib_ZX},
but this time, instead of measuring the population of the state $\protect\ket{00}$,
we measure the probability of populating the state $\protect\ket{0,-x}$.
This is achieved by applying a $R_{y}(\pi/2)$ rotation before measuring
the target qubit. Consequently, the outcome is sensitive to the ZY
term instead of the ZX term. Since PST suppresses the residual ZY
term and other coherent errors as well, we expect the calibration
curve to be flat as the amplitude stretch is varied. As shown in Fig.
\ref{fig: PST_calib_ZY}b, the experimental results validate this
prediction. In contrast, the $n=3$ curve without PST is not flat.
To see this more quantitatively, we calculate the $\sqrt{SSR}$ score
(SSR - Sum of Squared Residuals) that indicates the error in a fit
to a straight line, and find that it is ten times smaller with PST
compared to the same experiment without PST. For clarity, the fitted
straight lines are not shown. We used forty random twirling in this
experiment. }
\end{figure}

One might worry that for $\phi\neq0$ a residual ZY term will remain
and generate a coherent error. However as explained in Sec. \ref{subsec: PST-and-Calibration},
PST eliminates the small $a\sin\phi\,\sigma_{z}\otimes\sigma_{y}$
terms. To experimentally demonstrate this point we carry out a variant
of the previous experiment where a $\pi/2$ single-qubit $y$ rotation
is added to the target qubit just before the measurement. This makes
the survival probability sensitive to $ZY$ rather than to $ZX$.
Without PST, it is expected that the outcomes will show some dependence
on the amplitude since $\chi a\sin\phi\,\sigma_{z}\otimes\sigma_{y}$
is not exactly zero if $\phi\neq0$. This behavior is observed in
\ref{fig: PST_calib_ZY}a for different mitigation order $n$. However,
Fig. \ref{fig: PST_calib_ZY}b demonstrates that after applying PST
and KIK we find that the calibration curve is flat as expected. The
root of the sum square residuals $\sqrt{SSR}$ in fitting a straight
line to the data is presented in the yellow boxes. For $n=3$ the
$\sqrt{SSR}$ score with PST is an order of magnitude smaller compared
to the no PST case. In this last experiment we used the Taylor coefficients
in the KIK method in order to reduce the error bars. Furthermore,
for better visibility, in this experiment the sequence contained 41
repetitions of $\pi/2$ echo cross-resonance pulses. 

The error of the means $\sigma$ used for plotting the error bars
in the PST plot in Fig. 8 and Fig. 9 were calculated using 
\begin{equation}
\sigma^{2}=\frac{1}{n_{PST}-1}\sum_{j=1}^{n_{PST}}(R_{j}-\frac{1}{n_{PST}}\sum_{k=1}^{n_{PST}}R_{k})^{2},\label{eq: noisy means var}
\end{equation}
where $n_{PST}$ is the number of PST realizations used, and $R_{j}$
is the mitigated survival probability calculated by averaging over
all the shots in a given realizations. At first, it may seem counter
intuitive that this simple formula captures correctly the two components
of the variance: i) the variance in the measurement of each $R_{k}$
that  is determined by the number of shots in each realization, and
ii) the variance in the values of $R_{j}$ with respect to its mean
($\frac{1}{n_{PST}}\sum_{k=1}^{n_{PST}}R_{k}$) that is related to
the PST twirling of the coherent error. Such a separation to the variance
of means and means of variances is possible provided that a good estimation
of the means is known \citep{TotalVarianceBook}. However, our measurements
do not provide the exact means but the `noisy means' that includes
the additional uncertainty generated by the finite number of shots.
We numerically corroborate that the variance of noisy means formula
(\ref{eq: noisy means var}) agrees with the variance observed in
simulation. 

Finally we point out an important observation regarding the use of
virtual Z gates in the pseudo twirling Paulis. Due to their simplicity
and zero error rates, virtual Z gates \citep{VirtZ} are very popular
in quantum computers setups. In short, Z rotations can be pushed further
down the circuit by modifying the phases of pulses that generate gates
that do not commute with Z. However, despite their advantage, as explained
in Appendix V, a virtual Z gate is not suitable for generating the
Z gates needed for PST. Essentially, it makes twirlings that should
be different become the same, effectively reducing the space of available
twirlings. Thus, in our experiment, we implement the $Y_{pulse}$
gate by changing the phase of the $X$ pulse and not by implementing
$XZ_{v}$ (where $Z_{v}$ is a virtual Z gate). For $Z$ gates we
use $Z_{pulse}=Y_{pulse}X$ or $XY_{pulse}$. See also a similar effect
of virtual Z on dynamical decoupling experiments \citep{LidarCNOTdd}. 

\section{Summary and concluding remarks}

The working hypothesis underlying the present paper is that shorter
gates facilitate deeper circuits. However, the coherent errors in
short multi-qubit gates cannot be addressed with randomized compiling
(Pauli twirling) techniques as it hinges on the assumption that the
ideal gates are Clifford gates. Our analytical, numerical, and experimental
findings indicate that pseudo twirling can bridge this gap and successfully
mitigate coherent errors in non-Clifford gates. Pseudo twirling possesses
three advantages that separate it from the established randomized
compiling approach: i) Applicability to non-Clifford gates: it enables
the implementation of significantly shorter gates, resulting in lower
noise accumulation compared to Clifford-based implementations. ii)
Intra-gate coherent error suppression: pseudo twirling allows for
the suppression of coherent errors within gates and sliced pulses,
leading to substantial improvement in coherent error mitigation. iii)
Edges-PST: by applying pseudo twirling also at the edges of a circuit,
it is possible to suppress non-local multi-qubit coherent errors.
Moreover, the successful integration of PST with the Adaptive-KIK
error mitigation method leads to a substantial performance boost.
This boost can be instrumental in achieving quantum advantage in a
wide range of fields, including particle physics, condensed matter,
quantum chemistry, and more.
\begin{acknowledgments}
The authors acknowledge the use of IBM Quantum services for this work.
The views expressed are those of the authors, and do not reflect the
official policy or position of IBM or the IBM Quantum team. We thank
IAI/ELTA Systems for the assistance in running the experiments. Raam
Uzdin is grateful for support from the Israel Science Foundation (Grant
No. 2556/20). The support of the Israel Innovation Authority is greatly
appreciated.
\end{acknowledgments}

\bibliographystyle{apsrev4-2}
\bibliography{Refs_PST}

\section*{Appendix I - Quantum mechanics in Liouville space}

Within the conventional Hilbert space framework of Quantum Mechanics,
a quantum system of dimension $d$ is described by a density matrix
$\rho$ which has the dimensions $d\times d$. In addition, a general
completely positive quantum operation can be expressed as 
\begin{equation}
\rho'=\sum_{i}K_{i}\rho K_{i}^{\dagger},\label{eq:S1 Kraus representation}
\end{equation}
where $\rho'$ is the transformed density matrix and $\{K_{i}\}$
are Kraus operators that satisfying the relation $\sum_{i}K_{i}^{\dagger}K_{i}=I$
($I$ being the $d\times d$ identity matrix). In the Hilbert space
formalism, observables are described by a Hermitian operators $A$.
The expectation value of $A$ for system in the state $\rho$ reads
\begin{equation}
\left\langle A\right\rangle =\textrm{Tr}\left(A\rho\right).\label{eq:S2 expect value in Hilbert space}
\end{equation}

The Liouville space representation serves as an alternative mathematical
formulation that greatly aids in streamlining the expression of quantum
states and their transformations. In this model, the density matrix
is transformed into a state vector, indicated as $|\rho\rangle$,
with a dimensionality of $d^{2}\times1$, and a generic quantum transformations
$\mc O$ is described by a superoperator matrix with dimensions $d^{2}\times d^{2}$
such that 

\begin{equation}
|\rho'\rangle=\mathcal{\mathcal{O}}|\rho\rangle.\label{eq:S3 quantum operation in L space}
\end{equation}
Following the methodology of Ref. \citep{gyamfi2020fundamentals},
$|\rho\rangle$ is the column vector whose first $d$ components correspond
to the first row of $\rho$, the next $d$ components correspond to
the second row of $\rho$, and so forth. More explicitly, the vectorization
procedure of a $d\times d$ matrix $B$ reads $|B\rangle=\left(B_{11},B_{12},...,B_{1d},B_{21},B_{22},...,B_{2d},...,B_{d1},B_{d2},...,B_{dd}\right)^{\intercal}$,
where $B_{ij}$ is the $ij$ entry of $B$. With this convention,
in Liouville space a Kraus map (\ref{eq:S1 Kraus representation})
takes the form \citep{gyamfi2020fundamentals}
\begin{equation}
|\rho'\rangle=\sum_{i}K_{i}\otimes K_{i}^{\ast}|\rho\rangle,\label{eq:S4 Kraus in L space}
\end{equation}
where $K_{i}^{\ast}$ is the element-wise \emph{complex} conjugate
of $K_{i}$. Since unitary maps $\rho'=U\rho U^{\dagger}$ are a special
case of Kraus map , in Liouville space a unitary evolution is given
by 
\begin{equation}
|\rho'\rangle=\mathcal{U}|\rho\rangle=U\otimes U^{\ast}|\rho\rangle.
\end{equation}

Equation (\ref{eq:S4 Kraus in L space}) follows from the vectorization
rule for a product of three matrices $B$, $C$ and $D$ \citep{gyamfi2020fundamentals}
\begin{equation}
|BCD\rangle=B\otimes D^{\intercal}|C\rangle,\label{eq:S5 triple-prod identity}
\end{equation}
where the superscript $t$ denotes transposition (not Hermitian conjugate).
Equation (\ref{eq:S4 Kraus in L space}) follows by setting $B=K_{i}$,
$C=\rho$, and $D=K_{i}^{\dagger}$. 

Lastly, expectation value (\ref{eq:S2 expect value in Hilbert space})
of an observable matrix $A_{d\times d}$ can be neatly expressed in
Liouville space using the 'bra' vector $\bra A=\ket A^{\dagger}$.
That is, as a row vector $\langle A|=\left(A_{11}^{\ast},A_{12}^{\ast},...,A_{1d}^{\ast},...,A_{d1}^{\ast},A_{d2}^{\ast},...,A_{dd}^{\ast}\right)$.
Using the Hermiticity of $A$ we obtain that of the expectation value
of $A$ can be written as a 'bra-ket' inner product of $\ket A$ and
$\ket{\rho}$
\begin{align}
\langle A\rangle & =\sum_{i,j}A_{ji}\rho_{ij}\nonumber \\
 & =\sum_{i,j}A_{ij}^{*}\rho_{ij}\nonumber \\
 & =\langle A|\rho\rangle.
\end{align}

\section*{Appendix II - Derivation of the relation $\sum_{\alpha=1}^{2^{2n}}sgn(\alpha,\beta)\mathcal{P}{}_{\alpha}\mathcal{H}_{\gamma\protect\neq\beta}\mathcal{P}{}_{\alpha}=0$}

We remind that $sgn(\alpha,\beta)=tr[P_{\alpha}P_{\beta}P_{\alpha}P_{\beta}]/2^{n}$
is equal $+1$ ($-1$) if the two $n\text{-qubit}$ Paulis $P_{\alpha}$
and $P_{\beta}$ commute (anti commute). Starting by splitting the
sum into two terms.

\begin{align}
\sum_{\alpha=1}^{2^{2n}}sgn(\alpha,\beta)\mathcal{P}{}_{\alpha}\mathcal{H}_{\gamma\neq\beta}\mathcal{P}{}_{\alpha} & =\sum_{\alpha\in\beta^{+}}sgn(\alpha,\beta)\mathcal{P}{}_{\alpha}\mathcal{H}_{\gamma}\mathcal{P}{}_{\alpha}+\sum_{\alpha\in\beta^{-}}sgn(\alpha,\beta)\mathcal{P}{}_{\alpha}\mathcal{H}_{\gamma}\mathcal{P}{}_{\alpha},
\end{align}
where $\beta^{\pm}$ is the set of $\alpha$'s such that: $\mathcal{P}{}_{\alpha}\mathcal{H}_{\beta}\mathcal{P}{}_{\alpha}=\pm\mathcal{H}_{\beta}$
($sgn(\alpha,\beta)=\pm1$). Further simplification leas to

\begin{align}
\sum_{\alpha=1}^{2^{2n}}sgn(\alpha,\beta)\mathcal{P}{}_{\alpha}\mathcal{H}_{\gamma\neq\beta}\mathcal{P}{}_{\alpha} & =\sum_{\alpha\in\beta^{+}}\mathcal{P}{}_{\alpha}\mathcal{H}_{\gamma}\mathcal{P}{}_{\alpha}-\sum_{\alpha\in\beta^{-}}\mathcal{P}{}_{\alpha}\mathcal{H}_{\gamma}\mathcal{P}{}_{\alpha}\nonumber \\
 & =2\sum_{\alpha\in\beta^{+}}\mathcal{P}{}_{\alpha}\mathcal{H}_{\gamma}\mathcal{P}{}_{\alpha}.\label{eq: sum_alpha}
\end{align}
The last equality follows from the twirling property $\sum_{\alpha}\mathcal{P}{}_{\alpha}\mathcal{H}\mathcal{P}{}_{\alpha}=0$
that holds for any $\mc H$. Next, we define a two-element permutation
within $\beta^{+}$. To each element $\mathcal{P}{}_{\alpha_{1}^{+}}$
in $\beta^{+}$ we can assign another Pauli operator $\mathcal{P}{}_{\alpha_{2}^{+}}$
in $\beta^{+}$ using the transformation

\begin{equation}
\mathcal{P}{}_{\alpha_{2}^{+}}=\mathcal{P}{}_{\delta}\mathcal{P}{}_{\alpha_{1}^{+}},
\end{equation}
where $\mathcal{P}{}_{\delta}$ is chosen such $\mathcal{P}{}_{\delta}\mc H_{\beta}\mathcal{P}{}_{\delta}=\mc H_{\beta}$
and $\mathcal{P}{}_{\delta}\mc H_{\gamma}\mathcal{P}{}_{\delta}=-\mc H_{\gamma}$,
i.e. $[P_{\delta},P_{\beta}]=0$ and $\{P_{\delta},P_{\gamma}\}=0$.
To verify that $\alpha_{2}^{+}\in\beta^{+}$ we check if $\mathcal{P}{}_{\alpha_{2}^{+}}\mc H_{\beta}\mathcal{P}{}_{\alpha_{2}^{+}}=+\mc H_{\beta}$

\begin{equation}
\mathcal{P}{}_{\delta}\mathcal{P}{}_{\alpha_{1}^{+}}\mc H_{\beta}\mathcal{P}{}_{\delta}\mathcal{P}{}_{\alpha_{1}^{+}}=\mathcal{P}{}_{\alpha_{1}^{+}}\mathcal{P}{}_{\delta}\mc H_{\beta}\mathcal{P}{}_{\delta}\mathcal{P}{}_{\alpha_{1}^{+}}=\mathcal{P}{}_{\alpha_{1}^{+}}\mc H_{\beta}\mathcal{P}{}_{\alpha_{1}^{+}}=+\mathcal{P}{}_{\alpha_{1}^{+}}.
\end{equation}
which confirms that $\alpha_{2}^{+}\in\beta^{+}$. In the derivation
we use the fact that in Liouville space all $\mathcal{P}$ commute
with each other. We proceed by studying the outcome of $\mathcal{P}{}_{\alpha_{2}^{+}}\mc H_{\gamma}\mathcal{P}{}_{\alpha_{2}^{+}}$
compared to $\mathcal{P}{}_{\alpha_{1}^{+}}\mc H_{\gamma}\mathcal{P}{}_{\alpha_{1}^{+}}$

\begin{equation}
\mathcal{P}{}_{\alpha_{2}^{+}}\mc H_{\gamma}\mathcal{P}{}_{\alpha_{2}^{+}}=\mathcal{P}{}_{\delta}\mathcal{P}{}_{\alpha_{1}^{+}}\mc H_{\gamma}\mathcal{P}{}_{\delta}\mathcal{P}{}_{\alpha_{1}^{+}}=\mathcal{P}{}_{\alpha_{1}^{+}}\mathcal{P}{}_{\delta}\mc H_{\gamma}\mathcal{P}{}_{\delta}\mathcal{P}{}_{\alpha_{1}^{+}}=-\mathcal{P}{}_{\alpha_{1}^{+}}\mc H_{\gamma}\mathcal{P}{}_{\alpha_{1}^{+}}.
\end{equation}

We conclude that every element $\alpha\in\beta^{+}$ has counterpart
element that has opposite sign to the sum (\ref{eq: sum_alpha}).
As a result, the sum is zero. In the more general case $g_{\gamma}\mc H_{\gamma}\to\sum_{\gamma}g_{\gamma}\mc H_{\gamma}$
a different $P_{\delta}$ can be chosen to each element in the sum
so the proof above applies automatically to this case as well. Finally,
we point out that if it happens that $\{P_{\beta},P_{\gamma}\}=0$
it is possible to choose $P_{\delta}=P_{\beta}$.

\section*{Appendix III - derivation of $\protect\mc K_{pst}^{(m)}$ }

Under the assumption that the slices are sufficiently short that neither
the drive or coherent error change substantially along the slice we
can write 
\begin{equation}
\mc K_{pst}^{(m)}=e^{-idt\mc H(t_{1})}e^{+\mc L_{dt}^{pst}}e^{-idt\mc H(t_{2})}e^{+\mc L_{dt}^{pst}}.....e^{-idt\mc H(t_{m})}e^{+\mc L_{dt}^{pst}},
\end{equation}
 where $dt=T/m$, $t_{m}=T$, and $\mc L_{dt}^{pst}=dt^{2}\frac{1}{2^{2n}}\sum_{\alpha=1}^{2^{2n}}\mc L(P_{\alpha}H_{coh}P_{\alpha})$
(see Eq. (\ref{eq: LeffCoh})). By inserting the identities $\{U_{k}U_{k}^{\dagger}\}_{k=1}^{m}$
and $U_{m-k}^{\dagger}e^{-idt\mc H(t_{k+1})}=U_{m-k-1}^{\dagger}$
we obtain:
\begin{align}
\mc K_{pst} & =U_{m}[U_{m}^{\dagger}e^{-idt\mc H(t_{1})}e^{+\mc L_{dt}^{pst}}U_{m-1}][U_{m-1}^{\dagger}e^{-idt\mc H(t_{2})}e^{+\mc L_{dt}^{pst}}U_{m-2}]\nonumber \\
\times & [U_{m-2}^{\dagger}e^{-idt\mc H(t_{2})}e^{+\mc L_{dt}^{pst}}U_{m-3}][U_{m-3}^{\dagger}.....e^{-idt\mc H(t_{m})}e^{+\mc L_{dt}^{pst}}\nonumber \\
 & =U_{m}[U_{m-1}^{\dagger}e^{+\mc L_{dt}^{pst}}U_{m-1}][U_{m-2}^{\dagger}e^{+\mc L_{dt}^{pst}}U_{m-2}][U_{m-3}^{\dagger}e^{+\mc L_{dt}^{pst}}U_{m-3}]\nonumber \\
\times & U_{m-3}^{\dagger}.....U_{1}U_{1}^{\dagger}e^{-idt\mc H(t_{m})}e^{+\mc L_{dt}^{pst}}\nonumber \\
 & =U_{m}e^{U_{m-1}^{\dagger}\mc L_{dt}^{pst}U_{m-1}^{\dagger}}e^{U_{m-2}^{\dagger}\mc L_{dt}^{pst}U_{m-2}^{\dagger}}e^{U_{m-1}^{\dagger}\mc L_{dt}^{pst}U_{m-1}^{\dagger}}...e^{+\mc L_{dt}^{pst}}.
\end{align}
 By keeping the leading order in $\mc L_{dt}^{pst}$ we have
\begin{equation}
\mc K_{pst}=\mc U_{T}e^{\sum_{k}\mc U_{m-k}^{\dagger}\mc L_{dt}^{pst}\mc U_{m-k}}.
\end{equation}
In regular dissipation $\mc L_{dt}\propto dt$ so in the limit of
large m we obtain the standard first Magnus term. However the present
case $\mc L_{dt}^{PST}\propto dt^{2}=\frac{T}{m}dt$ and therefore
additional $T/m$ coefficient appears and we have
\begin{align}
\mc K_{pst}^{(m)} & =\mc U_{T}e^{\frac{T}{m}\sum_{k=0}^{m-1}\mc U_{k}^{\dagger}\frac{1}{2^{2n}}\sum_{\alpha=1}^{2^{2n}}\mc L(P_{\alpha}H_{coh}P_{\alpha})\mc U_{k}dt}\nonumber \\
 & \to\mc U_{T}e^{\frac{T}{m}\intop\mc U^{\dagger}(t)\frac{1}{2^{2n}}\sum_{\alpha=1}^{2^{2n}}\mc L(P_{\alpha}H_{coh}P_{\alpha})\mc U(t)dt}.
\end{align}

\section*{Appendix IV - using the operator norm for quantifying errors}

In our simulations we use the operator norm to describe the error
in the evolution operator. In this appendix we show that it is directly
related to the error in measuring observables. The operator norm of
a matrix $S$ is equal to its largest singular value. Yet, an alternative
definition is:
\begin{equation}
\nrm S_{op}=max_{\psi}\frac{|\braOket{\psi}{S^{\dagger}S}{\psi}|}{\sqrt{\braket{\psi}{\psi}}}.\label{eq: Sop}
\end{equation}
Applying the Cauchy--Schwarz inequality to the error in some expectation
value $\langle A|\rho_{f}\rangle=\langle A|\mc K|\rho_{0}\rangle$
with respect to the ideal value $\langle A|\mc U|\rho_{0}\rangle$
we have:
\begin{align}
\left|\braOket A{\mc K-\mc U}{\rho_{0}}\right| & \le\sqrt{\braket AA}\sqrt{\braOket{\rho_{0}}{(\mc K-\mc U)^{\dagger}(\mc K-\mc U)}{\rho_{0}}}\nonumber \\
 & \le\sqrt{\braket AA}\sqrt{\braket{\rho_{0}}{\rho_{0}}}\nrm{\mc K-\mc U}_{op},\label{eq: Op bound}
\end{align}
where we used (\ref{eq: Sop}) in the last step. Since the constants
$\sqrt{\braket AA}$ and $\sqrt{\braket{\rho_{0}}{\rho_{0}}}$ are
trivial to evaluate, we conclude that $\nrm{\mc K-\mc U}_{op}$ sets
a clear and easy to calculate bound on the error in any observable
$A$ and any initial state $\rho_{0}$. Notably, often the diamond
norm is used for comparing different maps. While the diamond norm
has some appealing features that can help in proofs, it is difficult
to evaluate it when the number of qubits increases. In contrast, the
operator norm is trivial to calculate. Furthermore, in the example
discussed in Fig. 5, we found that the operator norm and the diamond
norm deviate by less than ten percent without any features that distinct
one from each other.

Finally, we point out that (\ref{eq: Op bound}) can be slightly refined
when $\mc K$ is a trace-preserving map. For trace preserving maps
$\braOket I{\mc U}{\rho_{0}}=\braOket I{\mc K}{\rho_{0}}=1$ and therefore
$|\braOket A{\mc K-\mc U}{\rho_{0}}|=|\braOket{A+\alpha I}{\mc K-\mc U}{\rho_{0}}|$.
This implies that $\sqrt{\braket AA}$ can be replaced by $\sqrt{\braket{A+\alpha I}{A+\alpha I}}$.
Minimizing over $\alpha$ we find that 
\begin{align}
\left|\braOket A{\mc K-\mc U}{\rho_{0}}\right| & \le\sqrt{\braket{A'}{A'}}\sqrt{\braket{\rho_{0}}{\rho_{0}}}\nrm{\mc K-\mc U}_{op},\\
A' & =A-tr(A)I/tr(I).
\end{align}

\section*{Appendix V - the issue with using virtual gates in PST}

The basic idea behind the frame rotation used for virtual Z is as
follows. Consider the case where we need to implement an $R_{z}(\theta)$
gate followed by a transformation U. We can write it as: 

\begin{equation}
Ue^{-\frac{i}{2}\theta\sigma_{z}}\to e^{-\frac{i}{2}\theta\sigma_{z}}e^{+\frac{i}{2}\theta\sigma_{z}}Ue^{-\frac{i}{2}\theta\sigma_{z}}=e^{-\frac{i}{2}\theta\sigma_{z}}[e^{+\frac{i}{2}\theta\sigma_{z}}Ue^{-\frac{i}{2}\theta\sigma_{z}}].
\end{equation}
For simplicity, we assume that $U=e^{+i\phi P_{\alpha}}$ where $P_{\alpha}$
is a Pauli Matrix. If $[P_{\alpha},\sigma_{z}]=0$, $Ue^{-i\theta\sigma_{z}}=e^{-i\theta\sigma_{z}}U$
and the $e^{-i\theta\sigma_{z}}$ was pushed to the next layer. The
idea in virtual gate is to keep pushing the $R_{z}$ gates all the
way to the end of the circuit. Since the measurement is in the computational
basis, the final $R_{z}$ rotation has no impact.

Now let us consider the more interesting case where $P$ and $\sigma_{z}$
anti-commute:

\begin{align}
e^{+\frac{i}{2}\theta\sigma_{z}}Pe^{-\frac{i}{2}\theta\sigma_{z}} & =(\cos\frac{\theta}{2}+i\sin\frac{\theta}{2}\sigma_{z})P(\cos\frac{\theta}{2}-i\sin\frac{\theta}{2}\sigma_{z})\nonumber \\
 & =\cos^{2}\frac{\theta}{2}P+i\sin\frac{\theta}{2}[\sigma_{z},P]+\sin^{2}\frac{\theta}{2}\sigma_{z}P\sigma_{z}.
\end{align}
Using the anti-commutation:
\begin{equation}
(\cos^{2}\frac{\theta}{2}-\sin^{2}\frac{\theta}{2})P+i2\sin\frac{\theta}{2}\cos\frac{\theta}{2}\sigma_{z}P)=\cos\theta P+i\sin\theta\sigma_{z}P.
\end{equation}
Typically $P$ will be either $\sigma_{x}$ or $\sigma_{y}$. For
example, if it is $\sigma_{x}$ we get
\begin{equation}
e^{+\frac{i}{2}\theta\sigma_{z}}\sigma_{x}e^{-\frac{i}{2}\theta\sigma_{z}}=\cos\theta\sigma_{x}+i\sin\theta\sigma_{z}\sigma_{x}=\cos\theta\sigma_{x}+\sin\theta\sigma_{y},
\end{equation}
which means that the physical implementation of $e^{+\frac{i}{2}\theta\sigma_{z}}Pe^{-\frac{i}{2}\theta\sigma_{z}}$
just involves changing the phase of the pulse generating $P_{\alpha}$.
For example, an $R_{z}$ before the target qubit of ZX cross-resonance
pulse, will manifest in changing the phase of the cross-resonance
pulse.

Let us now consider how virtual Z work in PST
\begin{equation}
e^{-i\pi/4\mc H_{ZX}+i\mc H_{coh}}\to\mc P_{IZ}e^{+i\pi/4\mc H_{ZX}+i\mc H_{coh}}\mc P_{IZ}.
\end{equation}
In pseudo twirling, we use Pauli gates so $\theta=\pi$ in $R_{z}(\theta)$,
and therefore the virtualization will amount to changing the sign
of the drive without affecting the coherent error 
\begin{equation}
\mc P_{IZ_{v}}e^{+i\pi/4\mc H_{ZX}+i\mc H_{coh}}\mc P_{IZ_{v}}=e^{-i\pi/4\mc H_{ZX}+i\mc H_{coh}}\mc P_{IZ_{v}}\mc P_{IZ_{v}}=e^{-i\pi/4\mc H_{ZX}+i\mc H_{coh}}.
\end{equation}
 Thus, a virtual Z twirling is equivalent to the identity twirling
and will do nothing in this case. In contrast, when using physical
Z gates we get:

\begin{equation}
\mc P_{IZ}e^{+i\pi/4\mc H_{ZX}+i\mc H_{coh}}\mc P_{IZ}=e^{+\mc P_{IZ}(i\pi/4\mc H_{ZX}+i\mc H_{coh})\mc P_{IZ}}=e^{-i\pi/4\mc H_{ZX}+i\mc P_{IZ}\mc H_{coh}\mc P_{IZ}},
\end{equation}
which twirls the coherent error. Similarly, in IY twirling we aim
to generate
\begin{equation}
e^{-i\pi/4\mc H_{ZX}+i\mc H_{coh}}\to\mc P_{IY}e^{+i\pi/4\mc H_{ZX}+i\mc H_{coh}}\mc P_{IY}.
\end{equation}
However, if $\mc P_{IY}$ is generated via virtual Z: $\mc P_{IY}=\mc P_{IX}\mc P_{IZ_{v}}$
we obtain
\begin{align}
\mc P_{IX}\mc P_{IZ_{v}}e^{+i\pi/4\mc H_{ZX}+i\mc H_{coh}}\mc P_{IX}\mc P_{IZ_{v}} & =\nonumber \\
\mc P_{IX}e^{-i\pi/4\mc H_{ZX}+i\mc H_{coh}}\mc P_{IZ_{v}}\mc P_{IX}\mc P_{IZ_{v}} & =\nonumber \\
\mc P_{IX}e^{-i\pi/4\mc H_{ZX}+i\mc H_{coh}}\mc P_{IX}.
\end{align}
Thus, the IY twirling is indistinguishable from IX twirling. While
aiming to sample from the full set of twirlings, the use of virtual
Z and also Y using virtual Z, decimates the effective set of accessible
twirlings and consequently degrades the ability of PST to mitigate
the full scope of uncontrolled coherent errors.
\end{document}